\documentclass[12pt]{article}
\usepackage{geometry}
\usepackage[utf8]{inputenc}
\usepackage{graphics}
\usepackage{amsmath}
\usepackage{amssymb}
\usepackage{algorithmic}
\usepackage[colorlinks, linkcolor=red, anchorcolor=blue,citecolor=green]{hyperref}
\usepackage{algorithm}

\newtheorem{remark}{Remark}

\usepackage{mathrsfs}
\usepackage{natbib}
\usepackage{graphicx}
\usepackage{float} 
\usepackage{bbding}
\usepackage{color}
\usepackage{subfigure}
\usepackage{enumitem}
\usepackage[title]{appendix}
\usepackage{diagbox}
\begin{document}
\title{Fast Learning in Quantitative Finance with Extreme Learning Machine}
\author{Liexin Cheng\textsuperscript{a,b}, Xue Cheng\textsuperscript{a,b} \footnote{Corresponding author.
Email: chengxue@math.pku.edu.cn}, and Shuaiqiang Liu\textsuperscript{\S} \footnote{The views expressed in this paper are personal views of the author (SL) and do not necessarily reflect the views or policies of his current or past employers.}}

\maketitle
\vspace{-2em}
\small
\textit{
\begin{center}
    \textsuperscript{a} School of Mathematical Science, Peking University, Beijing, China\\
    \textsuperscript{b} Center for Statistical Science \& Key Laboratory of Mathematical Economics and Quantitative Finance, Peking University, Beijing, China \\
    \textsuperscript{\S} Delft Institute of Applied Mathematics, Delft University of Technology, The Netherlands \& ING Bank, Amsterdam, The Netherlands
\end{center}
}
\begin{abstract}

A critical factor in adopting machine learning for time-sensitive financial tasks is computational speed, including model training and inference. This paper demonstrates that a broad class of such problems, especially those previously addressed using deep neural networks, can be efficiently solved using single-layer neural networks without iterative gradient-based training. This is achieved through the extreme learning machine (ELM) framework. ELM utilizes a single-layer network with randomly initialized hidden nodes and output weights obtained via convex optimization, enabling rapid training and inference. We present various applications in both supervised and unsupervised learning settings, including option pricing, intraday return prediction, volatility surface fitting, and numerical solution of partial differential equations. Across these examples, ELM demonstrates notable improvements in computational efficiency while maintaining comparable accuracy and generalization compared to deep neural networks and classical machine learning methods. We also briefly discuss theoretical aspects of ELM implementation and its generalization capabilities. 

~
\end{abstract}

\section{Introduction}

Machine learning methods have become transformative tools in quantitative finance. In particular, artificial neural networks (ANNs),   especially those with multiple hidden layers or deep neural networks (DNNs), have achieved notable success in both mathematical model-based applications and data-driven applications. A comprehensive review of these developments on option pricing and hedging is provided by \cite{ruf2019neural}. More precisely, a prominent research direction involves using deep neural networks to accelerate computationally expensive tasks in model-based applications. Examples include  deep learning-based methods for calibrating stochastic volatility models (e.g., \cite{liu2019neural}  \cite{horvath2021deep} and the variant \cite{baschetti2024deep}), as well as  deep learning-based numerical schemes for solving Partial Differential Equations (PDEs), such as Physics-Informed Neural Networks (PINNs) \cite{raissi2019physics}, Deep BSDE methods for pricing European \cite{han2018solving} or American options \cite{chen2021deep}, among others (refer to the survey on machine learning methods for stochastic control \cite{hu2024recent}).  
Beyond model-based applications, deep neural networks  have also been widely used for processing large financial data, see the survey \cite{ozbayoglu2020deep}. These neural network-based data-driven methods can better handle challenging tasks over large datasets, e.g., high frequency return prediction and high dimensional financial data, compared to classical machine learning methods such as logistic regression (LR) and Gaussian process regression (GPR) (\cite{williams1995gaussian}).

From a practical point of view,  learning speed, in terms of both training and inference, is a critical factor for the adoption of DNN-based methods in finance.
Despite their expressive power,  DNN-based methods are often limited by the high computational costs associated with nonconvex optimization and deep architectures. Training DNNs typically involves computationally intensive  optimization using iterative gradient-based algorithms such as stochastic gradient descent (SGD). These challenges hinder their deployment in real-time environments or time sensitive applications that require rapid model training and inference. 
In this work, we demonstrate that a broad class of such problems can be addressed more efficiently using random feature neural networks, specifically through the framework of Extreme Learning Machines (ELM).

Introduced by \cite{huang2006extreme}, ELM employs a single-hidden-layer architecture in which the hidden weights are randomly initialized and the output weights are determined via linear least squares. This framework transforms training neural networks into a convex optimization problem, which guarantees convergence while eliminating the need for iterative backpropagation. On the one hand, this leads to a faster training process while maintaining competitive accuracy. 
On the other hand, ELM offers higher inference speed, as single-hidden-layer networks can be efficiently implemented (e.g., via parallel computing). In contrast, deep neural networks employ a sequential structure, where each layer must wait for the output of the preceding layer.  Moreover, the mathematical foundations of ELM have been rigorously studied, including its universal approximation capabilities (\cite{huang2006universal, neufeld2023universal}), and error bounds for approximating sufficiently regular functions \cite{gonon2023approximation}.

 Although ELM has seen success in various fields as reviewed in \cite{wang2022review}, its potential in financial applications remains underexplored.  The fast learning capability makes ELM particularly well-suited for time-sensitive financial tasks. We will investigate  the application of ELM to representative supervised and unsupervised learning tasks in finance.

In supervised learning, the process typically involves two key phases: an initial offline training stage,  followed by frequent online inference operations. In this setting, the speed of model inference is  critical ,  for instance,  in calibrating  stochastic volatility models using deep learning techniques (e.g., \cite{horvath2021deep}, \cite{liu2019neural}). Thanks to its single-hidden-layer structure, ELM often outperforms DNNs in terms of model inference speed.    Moreover, ELM exhibits strong generalization capabilities during the model inference stage, which help mitigate overfitting in data-driven tasks. This advantage is especially evident when compared to classical methods such as Gaussian process regression for implied volatility surface fitting \cite{de2018machine}, and logistic regression  for high-frequency return prediction, as will be demonstrated in this paper.

While supervised learning tasks may not inherently require rapid training, since the training process  is often  done once and offline, there are many scenarios where faster training remains beneficial. For example, the  parametric pricing functions approximated by DNNs in \cite{horvath2021deep} and \cite{liu2019neural} are constrained by the model parameter bounds defined in the training phase. During model calibration, if market option quotes fall outside these predefined ranges, the DNN must be retrained to accommodate the out-of-range values.  In such scenarios, ELM can swiftly adapt to expanded parameter bounds without suffering from slow retraining.   Similarly, in applications where model inference must continuously incorporate new data (e.g., forecasting high-frequency financial time series), ELM can offer a practical advantage through their ability to perform rapid and frequent retraining. 

In unsupervised learning, we focus on solving PDEs by means of neural networks, as PDEs are  widely utilized in finance. Notable examples include the well-known Black–Scholes PDE for option pricing \cite{black1973pricing}, PDEs for derivative valuation adjustments (XVA), which gained prominence after the 2007–2008 financial crisis \cite{burgard2011partial}, among others.
Traditional numerical methods, such as finite difference, finite element, and finite volume schemes, require discretizing the computational domain. 
However, as the dimensionality of the PDE increases (e.g., in the case of options on multiple underlying assets),  these methods become computationally infeasible. This is due to the exponential growth in grid points, a phenomenon known as the curse of dimensionality. To address this challenge, neural networks offer a powerful tool for solving high-dimensional PDEs. 

One of such deep learning-based numerical methods are PINNs, which approximate the solution by minimizing a composite loss function that consists of both the PDE residual and the initial/boundary conditions. For example,  \cite{salvador2020financial} applies PINNs  to the pricing of European and American options.  However, classical PINNs often suffer from slow training due to the reliance on iterative gradient-based optimization. In this context, ELM provides a promising alternative to accelerate the  process of training neural networks. Recent work \cite{gonon2023random} theoretically proves that ELMs can learn solutions to certain classes of financial PDEs (e.g., Black–Scholes-type equations) without  the curse of dimensionality. 
Although \cite{gonon2023random} performs the numerical experiments under a supervised learning framework based on known input–output data, it nonetheless highlights the potential of ELM approximating the solution to high-dimensional financial PDEs. 
Furthermore, ELM has also been successfully applied as a substitute for deep neural networks in solving path-dependent PDEs arising in rough volatility models \cite{jacquier2023random}. 
Building on these advantages, our work advances the field by integrating ELM into the PINN framework for solving financial PDEs in a fast and efficient way.

The remainder of this paper is organized as follows. Section \ref{sec 2} outlines the theoretical foundations of ELM, including the original ELM and its enhanced variants. Section \ref{sec 3} explores supervised learning applications, focusing on inference speed, incremental training, and comparative performance against DNNs, GPR and LR. Section \ref{sec 4} discusses unsupervised learning via ELM-based PINNs, emphasizing their role in solving multidimensional financial PDEs. Section \ref{sec 5} concludes.

\section{Theory of ELM}
\label{sec 2}
\subsection{ELM}
An ELM is a single hidden layer feed-forward neural network whose hidden-layer weights are randomly generated before the training process. ELM can be trained quite efficiently for both classification and regression problems. The training of ELM solves a least-square problem instead of gradient-descent algorithms in traditional neural networks. The following summarizes:

Given a set of training data $(\mathbf{x}_j, y_j), j = 1, 2, \cdots, N$, where $\mathbf{x}_j \in \mathbf{R}^d$ and $y_j \in \mathbf{R}$, the output function of ELM is 
\begin{equation}
    \label{ELM}
    \text{ELM}(\mathbf{x}_j):= f_L(\mathbf{x}_j) =  \sum_{i=1}^L \beta_i G(\mathbf{w}_i \cdot \mathbf{x}_j + b_i), \quad j = 1, 2, \cdots, N, 
\end{equation}
where $f_L: \mathbf{R}^d \to \mathbf{R}$, $L$ is the number of hidden neurons, $G(\cdot)$ is an activation function, $\mathbf{w}_i \in \mathbf{R}^d$  is the weight vector that connects the input layer to the $i$th hidden node and $b_i \in \mathbf{R}$ is the bias of the hidden node. The weights and biases $(\mathbf{w}_i, b_i)$ are generated independently and randomly from a continuous distribution $D \in \mathbf{R}^{L + 1}$. We also write the $N$ equations \eqref{ELM} compactly as:
\begin{equation}
    \label{ELM compact}
    \mathbf{H} \cdot \boldsymbol{\beta} = \mathbf{Y},
\end{equation}
where $ \mathbf{H} \in \mathbf{R}^{N \times L}$ is the hidden layer output matrix with the $(j, i)$ term $G(\mathbf{w}_i \cdot \mathbf{x}_j + b_j)$, $\boldsymbol{\beta} = [\beta_1, \dots, \beta_L]^\top$ and $\mathbf{Y} = [y_1, \dots, y_N]^\top.$

The least square solution of the linear equation system is
$$\boldsymbol{\beta}=\mathbf{H}^{\dagger} \mathbf{Y},$$
where $\mathbf{H}^{\dagger}$ is the generalized Moore-Penrose inverse of matrix $\mathbf{H}$. It has been show in \cite{huang2006extreme} that $H$ is of full column rank with probability one, under mild conditions of the neural network. Then, if $L \le N$, we take $\mathbf{H}^{\dagger} 
= (\mathbf{H}^{\top} \mathbf{H})^{-1}\mathbf{H}^{\top},$ 
in which case a Cholesky decomposition with complexity $O(L^3)$ can be applied to solve $\mathbf{\beta}$. In more general cases where $H$ is rank deficient, the singular value decomposition can be used with numerical stability. Moreover, for large sparse situations, iterative methods such as conjugate gradient or GMRES can be more efficient.

It was also shown in \cite{huang2006universal} that given any positive value $\epsilon > 0$, we can find an $L \le N$ such that 
$\|\mathbf{H}\mathbf{\beta} - \mathbf{Y}\|_2 < \epsilon$
under mild network structure assumptions. One can also add a regularization term in the solution by minimizing $\|\mathbf{H} \cdot \boldsymbol{\beta} - \mathbf{Y}\|_2^2 + C \|\boldsymbol{\beta}\|_2^2$, which results in 
\begin{equation}
    \boldsymbol{\beta} = (\mathbf{H}^{\top} \mathbf{H} + C\mathbf{I})^{-1}\mathbf{H}^{\top} \mathbf{Y},
\end{equation}
where $C$ is a constant that determines the degree of regularization. Unlike conventional feedforward networks, ELM results in a convex optimization problem.

\subsection{Incremental ELM}
Following the baseline ELM (cf. \cite{huang2006extreme}), incremental ELM methods were proposed to automatically determine network architectures given an application scenario. I-ELM by \cite{huang2006universal} adds randomly generated nodes to the hidden layer one by one and fixes the output weights of the existing hidden nodes when a new hidden node was added. As an improvement to I-ELM, CI-ELM (\cite{huang2007convex}) recalculated the entire output weights every time a new hidden node is added, which yields faster error convergence in various datasets while keeping the simplicity of I-ELM. That is, the reduction in testing RMSE as the number of hidden nodes increases becomes faster in CI-ELM than in I-ELM. Later, \cite{feng2009error} proposed EM-ELM that increases nodes one by one or group by group based on error-minimization criteria. EM-ELM was shown to achieve even faster convergence rates and reduced computational complexity. 

In \cite{xu2016incremental}, an improved EM-ELM algorithm, called Enhanced Incremental Regularized ELM (EIR-ELM), was developed. Compared with EM-ELM, the new algorithm achieved better generalization performance by introducing regularization method and by selecting hidden nodes to be added to the network. The implementation of the EIR-ELM algorithm is summarized below.

For a given training data set $\aleph=\left\{\left(\mathbf{x}_i, \mathbf{t}_i\right) \mid \mathbf{x}_i \in \mathbf{R}^n, \mathbf{t}_i \in \mathbf{R}^m, i=\right.$ $1, \ldots, N\}$, the initial number of hidden nodes $N_0$, the maximum number of hidden nodes $N_{\max }$ and the expected learning accuracy $\epsilon$ :

~\\I. Initialize the neural network:
\begin{enumerate}[itemsep=1pt,parsep=2pt,topsep=2pt]
    \item Assign the input weights vectors $\mathbf{w}_i$ and basis $b_i, i=1, \ldots, N_0$ randomly.
    \item Calculate the hidden layer output matrix $\mathbf{H}_0$.
    \item Calculate the output weight $\boldsymbol{\beta}_0 . \boldsymbol{\beta}_0=\mathbf{D}_0 \mathbf{T} :=\left(\mathbf{H}_0^T \mathbf{H}_0+\mathbf{C I}\right)^{-1}$ $\mathbf{H}_0^T \mathbf{T}$, where $\mathbf{T}=\left[\mathbf{t}_1, \ldots, \mathbf{t}_{\mathrm{N}}\right]^{\mathrm{T}}$.
    \item Let $s=N_0$, calculate the learning accuracy $\epsilon_s$.
\end{enumerate}

~\\II. Update the network recursively. While $s<N_{\max }$ and $\epsilon_s<\epsilon$
\begin{enumerate}[itemsep=1pt,parsep=2pt,topsep=2pt]
    \item Let $s=s+1$.
    \item For $i=1: \mathrm{k}$
    \begin{enumerate}[itemsep=1pt,parsep=2pt,topsep=2pt]
        \item Generate a new hidden node $i$ randomly, $\mathbf{H}_s^i=\left[\mathbf{H}_{s-1}, \mathbf{v}_s^i\right]$
        \item Update the output weight as follows:

$$
\begin{aligned}
& \mathbf{M}_s=\frac{\mathbf{v}_s^{i T}\left(\mathbf{I}-\mathbf{H}_{s-1} \mathbf{D}_{s-1}\right)}{\mathbf{v}_s^{i T}\left(\mathbf{I}-\mathbf{H}_{s-1} \mathbf{D}_{s-1}\right) \mathbf{v}_s^i+\mathbf{C}} \\
& \mathbf{L}_s=\mathbf{D}_{s-1}\left(\mathbf{I}-\mathbf{v}_s^i \mathbf{M}_s\right), \quad \boldsymbol{\beta}_s^i=\mathbf{D}_s^i \mathbf{T}=\left[\begin{array}{c}
\mathbf{L}_s \\
\mathbf{M}_s
\end{array}\right] \mathbf{T}
\end{aligned}
$$
\item Calculate the cost function $J 
= \left\|\mathbf{H}_s^i \boldsymbol{\beta}_s^i-\mathbf{T}\right\|^2+C\left\|\boldsymbol{\beta}_s^i\right\|^2$.
    \end{enumerate}
    \item Choose the hidden node $k$ that has the smallest cost function then: $\mathbf{H}_s=\mathbf{H}_s^k, \mathbf{D}_s=\mathbf{D}_s^k$ and $\boldsymbol{\beta}_s=\boldsymbol{\beta}_s^k$.
    \item Calculate the new learning accuracy $\epsilon_s$.
\end{enumerate}
\section{Applications in Supervised Learning}
\label{sec 3}
In this section, we apply ELM and its variants to three supervised learning tasks. In the subsequent applications, a normal distribution with zero mean is adopted as the sampling distribution of the hidden nodes. The standard deviation, commonly referred to as the scale parameter in this context, is treated as an undetermined hyperparameter requiring empirical specification.

\subsection{Learning Parametric Pricing Functions}
We employ ELM to approximate the IVS function under the Heston and rough Heston stochastic volatility models. The framework processes inputs $(\Theta, T, k)$, where $\Theta$ denotes the set of model parameters, $T$ represents the option's time to maturity, and $k \equiv \frac{K}{F_0}$ represents moneyness(where $K$ is the strike price and $F_0$ denotes the current forward price). In DNN-based learning methods, the parametric pricing functions which DNN has learned are constrained by the model parameter bounds defined during the training phase.
In other words, the model parameters are confined to prespecified lower and upper bounds.
If market data falls outside this range (e.g., time to maturity exceeds the trained upper limit defined during the training dataset),
the DNN may require retraining to accommodate the new values. In contrast, ELMs offer an
efficient alternative that facilitates rapid adaptation to expanded
parameter domains without suffering from slow retraining.

We generate implied volatility from numerical solutions using the COS method (\cite{fang2009novel}), which yields a dataset of 100,000 samples. Each input vector comprises seven (eight) dimensions: $(k, T, \rho, \kappa, \sigma, \theta, v_0)$ for Heston ($(k, T, \rho, \alpha, \gamma, \nu, \theta, v_0)$ for rough Heston), where $k$ spans $[0.714, 1.667]$ for Heston and $[0.6, 1.4]$ for rough Heston, and $T$ ranges from $0.10$ to $3.00$ years for Heston and from $0.05$ to $3.00$ years for rough Heston. Consider a risk-neutral filtered probability space $(\mathbb{Q}, \Omega, \mathcal{F}, \{\mathcal{F}_t\}_{t\ge 0})$, $(\rho, \kappa, \sigma, \theta, v_0)$ correspond to Heston model parameters governing correlation, mean reversion rate, volatility of volatility, long-term variance, and initial variance, respectively:
  $$
  \left\{ \begin{aligned}
   &\mathrm{d}S_t/ S_t = r\mathrm{d}t + \sqrt{v_t}\mathrm{d}W_t\\
   &\mathrm{d}v_t = \kappa (\theta - v_t)\mathrm{d}t + \sigma\sqrt{v_t}\mathrm{d}Z_t,
   \end{aligned}\right.$$
   where $\mathrm{d}W_t\mathrm{d}Z_t = \rho \mathrm{d}t$ with $\mathbb{Q}$-Brownian motions $W$ and $Z$. Furthermore, $\rho, \alpha, \gamma, \nu, \theta, v_0$ correspond to rough Heston model parameters that govern correlation, roughness, mean reversion rate, volatility of volatility, long-term variance, and initial variance, respectively:
\begin{equation}
\left\{\begin{aligned}
&\mathrm{d} S_t / S_t = r\mathrm{d}t + \sqrt{v_t} \mathrm{d} W_t, \\
&v_t =v_0+\frac{1}{\Gamma(\alpha)} \int_0^t(t-s)^{\alpha-1} \gamma\left(\theta-v_s\right) \mathrm{d} s+\frac{1}{\Gamma(\alpha)} \int_0^t(t-s)^{\alpha-1} \gamma \nu \sqrt{v_s} \mathrm{d} Z_{s},\\
\end{aligned}\right.
\end{equation}
where $\mathrm{d}W_t\mathrm{d}Z_{t} = \rho \mathrm{d}t$ as before. 

The parameter ranges of Heston model are $\rho \in (-1, 0), \kappa \in (0, 4), \sigma \in (0, 0.5), \theta \in (0, 0.1), v_0 \in (0, 0.5)$, and those of rough Heston model are $\rho \in (-0.9, 0), \alpha \in (0.55, 0.95), \gamma \in (0.01, 3.0), \nu \in (0.01, 1.0), \theta \in (0.15, 0.5), v_0 \in (0.5, 1.0)$. The output implied volatility values exhibit a mean of 0.276 (0.230) and a standard deviation of 0.177 (0.141) for the Heston (rough Heston) model. The dataset is partitioned into two subsets: 80,000 training samples and 20,000 test samples. 

\begin{figure}[!ht]
	\centering
	\begin{minipage}{0.49\linewidth}
\vspace{3pt}\centerline{\includegraphics[width=\textwidth]{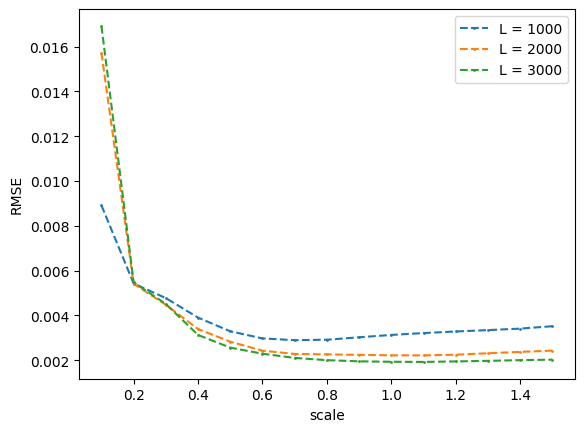}}
		\centerline{}
	\end{minipage}
	\begin{minipage}{0.49\linewidth}
		\vspace{3pt}
\centerline{\includegraphics[width=\textwidth]{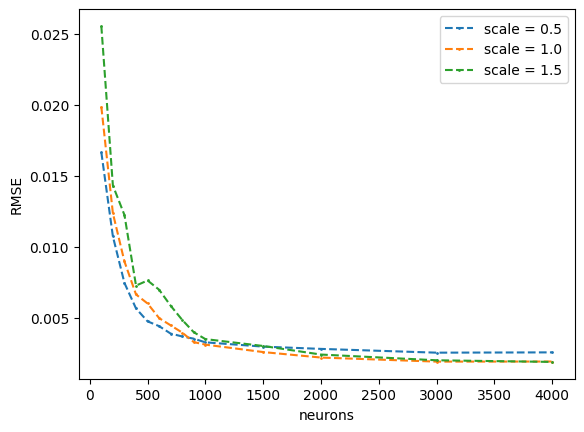}}
	 
		\centerline{}
	\end{minipage}
	\caption{Performance under different network parameters on the testing dataset. The left subfigure illustrates the variation in RMSE across the dataset with respect to the scale parameter of the sampling distribution, while the right subfigure demonstrates the relationship between RMSE and the number of hidden nodes in the network. A sine activation function is employed throughout the analysis, with all experiments conducted using a sample size of 100,000.}
	\label{scale}
\end{figure}

We first examine the dependence of the ELM performance on the network sampling distribution's scale parameter and the hidden node number. Figure \ref{scale} demonstrates these relationships: the left panel illustrates RMSE variation in the test sample with respect to the distribution scale, while the right panel quantifies test-sample RMSE as a function of node number. The error converges as neurons increase. The scale parameter selection, however, reveals an inherent trade-off. Excessively small scale values make the network too restrictive by limiting the diversity of features generated through weight and bias configurations. Conversely, excessively large scale parameters induce sparsity in the feature representations produced during training, which requires a larger node number to maintain approximation accuracy. An well-learned pricing function therefore requires balancing the node number with scale parameters.
    
\begin{table}[!ht]
    \centering
        \caption{Comparison of ELM and GPR performance on Heston and rough Heston models. ELM configurations with 1000 and 3000 hidden nodes are shown. A sine activation is employed for ELM. Both GPR and ELM are performed without tuning hyperparameters. The sample is randomly selected from the original data with size 15,000 for both models. Indicator MAE stands for mean absolute error and indicator MAPE stands for mean absolute percentage error calculated as $|(y_{\text{pred}} - y_{\text{true}}) / y_{\text{true}}| \times 100 \%$.} 
    \begin{tabular}{ccccc}
         & CPU Time (s) & RMSE & MAE\footnote{MAE stands for mean absolute error calculated as $|y_{\text{pred}} - y_{\text{true}}|$.} & MAPE (\%)\footnote{MAPE stands for mean absolute percentage error calculated as $\left|\frac{y_{\text{pred}} - y_{\text{true}}}{y_{\text{true}}}\right| \times 100 \%$.} \\\hline
  \multicolumn{5}{c}{Panel A: Training Process}\\
   \multicolumn{5}{l}{Panel A.1: Heston Model}\\\hline
       ELM (1000) & 0.48 & 0.00235 & 0.00163 & 3.98\\
       ELM (3000) & 2.23 & 0.00090 & 0.00062 & 1.54\\
      GPR & 14.71 & $9.66\times 10^{-6}$ &  $5.45\times 10^{-6}$ & 0.13\\
     \multicolumn{5}{l}{Panel A.2: Rough Heston Model}\\\hline
     ELM (1000) & 0.73 & 0.00177 & 0.00116 & 3.05\\
       ELM (3000) & 2.53 & 0.00077 & 0.00051 & 1.23\\
      GPR & 15.50 & $7.36\times 10^{-7}$ &  $4.11\times 10^{-7}$ & 0.02 \\\hline
    \multicolumn{5}{c}{Panel B: Testing Process}
         \\
    \multicolumn{5}{l}{Panel B.1: Heston Model}\\\hline
       ELM (1000) & 0.05 & 0.00292 & 0.00190 & 5.28\\
       ELM (3000) & 0.15 & 0.00173 & 0.00102 & 3.27\\
      GPR & 0.27 & 0.00243 & 0.00129 & 5.13\\
       \multicolumn{5}{l}{Panel B.2: Rough Heston Model}\\\hline
           ELM (1000) & 0.06 & 0.00201 & 0.00130 & 4.32\\
       ELM (3000) & 0.15 & 0.00135 & 0.00080 & 2.96\\
      GPR & 0.26 & 0.00160 & 0.00060 & 1.58\\\hline
    \end{tabular}
    \label{Com1}
\end{table}


To empirically demonstrate ELM's comparative advantages, we compare the small-sample (under size 20,000) performance of ELM with GPR method. A brief introduction of GPR can be found in Appendix \ref{GPR}. The GPR implementation aligns with the specifications outlined in \cite{de2018machine}.

\begin{figure}[!ht]
	\centering
	\begin{minipage}{0.49\linewidth}
\vspace{3pt}\centerline{\includegraphics[width=\textwidth]{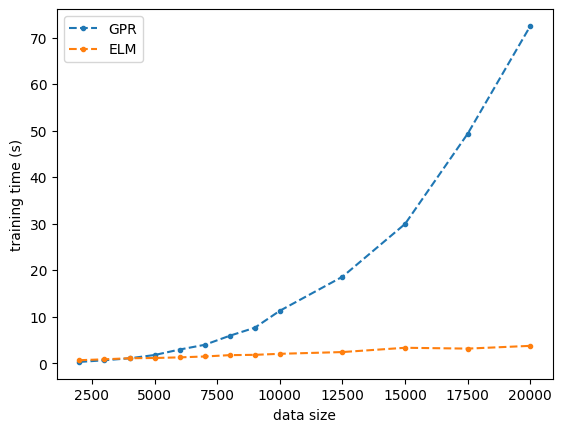}}
		\centerline{}
	\end{minipage}
	\begin{minipage}{0.49\linewidth}
		\vspace{3pt}
\centerline{\includegraphics[width=\textwidth]{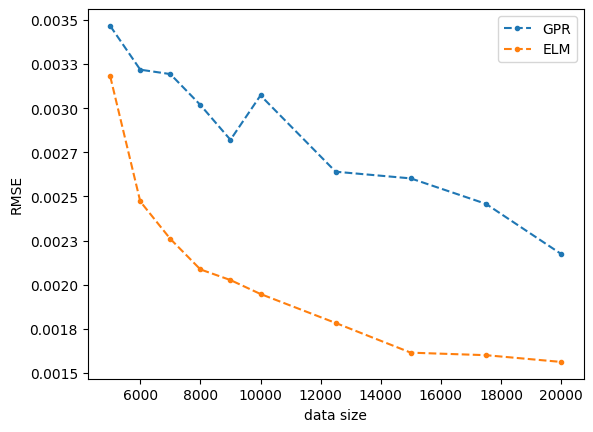}}
	 
		\centerline{}
	\end{minipage}
	\centering
	\begin{minipage}{0.49\linewidth}
\vspace{3pt}\centerline{\includegraphics[width=\textwidth]{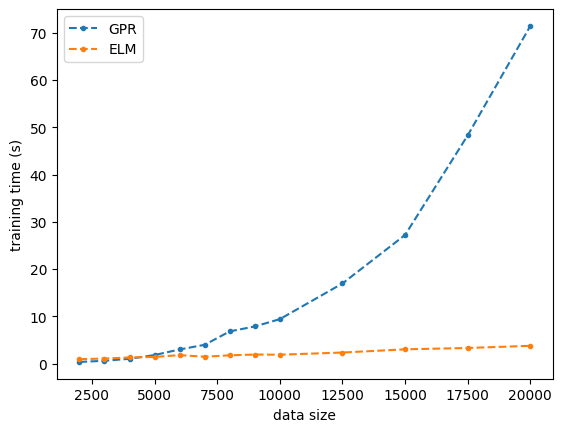}}
		\centerline{}
	\end{minipage}
	\begin{minipage}{0.49\linewidth}
		\vspace{3pt}
\centerline{\includegraphics[width=\textwidth]{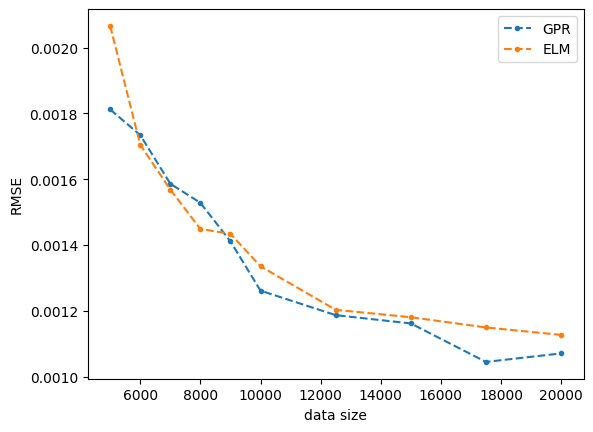}}
	 
		\centerline{}
	\end{minipage}
	\caption{Comparison of training speed between ELM ($L = 3000$) and GPR in the tasks of learning Heston pricing model (first row) and rough Heston pricing model (second row). Both GPR and ELM are performed without tuning hyperparameters. The left panel shows the increase in training time with respect to data size, and the right panel shows the corresponding out-of-sample RMSE in a testing sample size $3000$.}
	\label{time}
\end{figure}

We first compare the training process between ELM and GPR. As evidenced in Table \ref{Com1}, an ELM architecture with 3,000 hidden nodes achieves comparable predictive accuracy while avoiding the overfitting suffered by GPR, whose training error is significantly smaller than the test error. In the aspect of training, GPR requires constructing an $N\times N$ kernel matrix that results in a computational complexity scaling fast with data size. In comparison, ELM's computational complexity mainly scales with the neuron number under the Cholesky decomposition, resulting in several times faster training speed. Figure \ref{time} further elucidates this disparity through the comparison of training time and testing RMSE of Heston model (first row) and rough Heston model (second row): the right column demonstrates at least comparable performance metrics, while the left column shows that ELM’s computational time scales steadily and slowly with dataset size. In contrast, GPR exhibits rapid escalation in computational cost.

	 
\begin{figure}[!ht]
	\centering
	\begin{minipage}{0.49\linewidth}
\vspace{3pt}\centerline{\includegraphics[width=\textwidth]{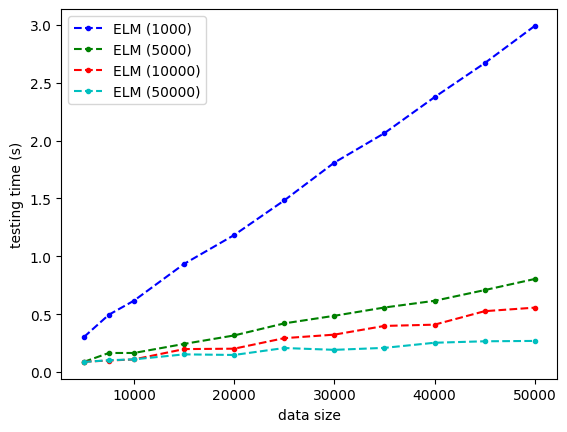}}
		\centerline{}
	\end{minipage}
    \begin{minipage}{0.49\linewidth}
		\vspace{3pt}
\centerline{\includegraphics[width=\textwidth]{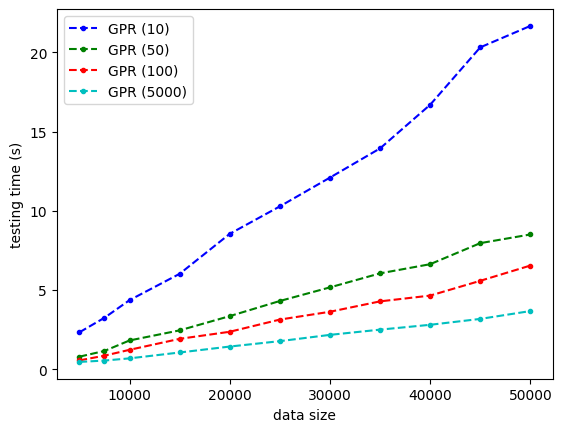}}
	 
		\centerline{}
	\end{minipage}

    \begin{minipage}{0.49\linewidth}
		\vspace{3pt}
\centerline{\includegraphics[width=\textwidth]{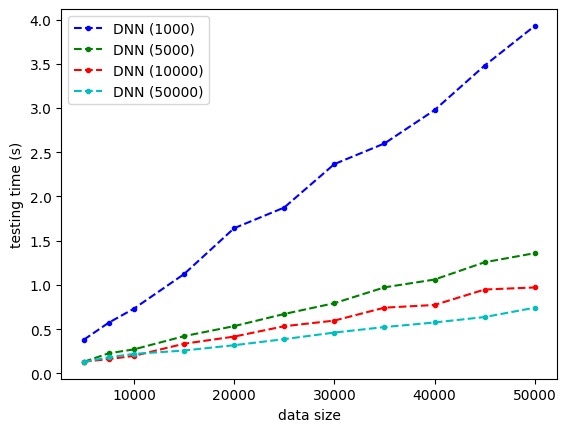}}
	\end{minipage}
	\begin{minipage}{0.49\linewidth}
		\vspace{3pt}
\centerline{\includegraphics[width=\textwidth]{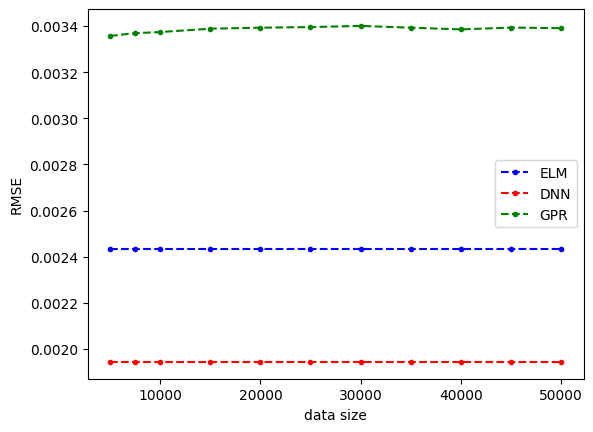}}
	\end{minipage}
	\caption{Comparison of the inference speed between ELM ($L = 3000$) , GPR and DNN (four hidden layers with $L = 750$ in each layer). A training sample of 12,000 is used. And the testing procedure is performed on a 3000 sample size under different batch sizes in order to avoid the memory issue}.
	\label{test time}
\end{figure}

We then compare the inference speed between ELM, GPR and DNN. Deep neural networks are needed for the problems of deep calibration or deep option pricing. We intend to show the efficiency of a single-layer network like ELM in the inference procedure. To this end, a DNN with a commonly used four-layer structure, each with 750 neurons, is incorporated. Note that the total number of neurons in DNN is the same in ELM. Figure \ref{test time} evaluates the CPU time requirements of inference in varying batch sizes, where the test sample is divided into batches of specified size, and the testing time represents the cumulative execution duration across all batches. Experimental results indicate that, with comparable testing performances, ELM achieves superior inference speed compared to GPR and DNN: under a datasize of 50,000 and optimized batch sizes, ELM takes 0.25 seconds and is around 15 (4) times faster than GPR (DNN), respectively.

As also shown in Figure \ref{test time}, the batch size of ELM/DNN can be significantly larger than that of GPR. This is because the dimensionality of the kernel matrix of GPR grows linearly with data size, leading to a computational complexity of $O(N^3)$ and spatial complexity $O(N^2)$, making it impractical for large datasets ($N > 10 ^ 5$). ELM circumvents this constraint through fixed-dimensional computations governed by the hidden node number (e.g. Cholesky decomposition with $O(L^3)$ computational complexity and spatial complexity increasing linearly with data size), which ensures both training efficiency and low memory requirements to large-scale tasks.

In summary, we find that with learning accuracy resembling GPR, ELM can train as well as infer much faster. In particular, the inference speed is important in practice, and ELM can be applied to large-scale datasets with better efficiency and memory requirements. In addition, a multi-layer DNN with the same number of neurons infer slower than ELM, making ELM a possible alternative for the deep calibration problem. Finally, we also examine how the EIR-ELM  streamlines network architecture through strategic node selection, see Appendix \ref{EIR-Simp} for details.

\subsection{Return Prediction}
This study implements ELM and LR to forecast directional movements in intraday stock prices. The dataset comprises high-frequency price data and order book records for Ping An Bank (000001.XSHE) from January 2, 2020 to December 31, 2022. We define the binary target variable \( y(i) \) as:  
\[
y(i) = 
\begin{cases} 
+1 & \text{if } S_{t_{i} + \Delta t} \geq S_{t_i} \quad  \\ 
-1 & \text{otherwise}, 
\end{cases}
\]  
where \( \Delta t = 5 \) minutes, and \( S_t \) denotes the asset price at time \( t \). This encodes the direction of 5-minute price changes for binary classification.   Twelve predictive features, derived from tick-level transaction histories and limit order book dynamics (detailed in Appendix \ref{A}), constitute the input vector.

Of the twelve quarters in the dataset, we consecutively select eight quarters for the evaluation of ELM and LR, with a total of three consecutive periods: 2020-01-02 $\sim$ 2021-12-31, 2020-07-01 $\sim$ 2022-06-30, 2021-01-04 $\sim$ 2022-12-30. For each window, models are initially trained on the first six quarters. During the subsequent two-quarter testing phase, daily recalibration is performed by incrementally integrating the prior trading day’s data, ensuring continuous adaptation to evolving market conditions.

\begin{table}[H]
    \centering
    \caption{Performance comparison between ELM and LR models. For each training/test set, the first value represents classification accuracy (\%) and the second value shows the F1-score (\%). Compared with classification accuracy, F1-score takes into account the imbalance in the class distribution.} ELM configurations used 30/300 hidden nodes with sine activation and simulation $\mathrm{scale}=0.01$. LR was trained for 1000 iterations to ensure convergence.
    \begin{tabular}{lccccc}
    & \multicolumn{2}{c}{Training} & \multicolumn{2}{c}{Test} & CPU Time (s) \\
    & Accuracy & F1-score & Accuracy & F1-score & \\ \hline
    \multicolumn{6}{l}{Period 1} \\ \hline
    ELM (30) & 55.69 & 47.98 & 54.65 & 45.92 & \textbf{1.77} \\
    ELM (300) & 54.98 & 53.02 & 53.47 & 50.42 & 24.98 \\
    LR & 55.53 & 45.60 & 54.98 & 43.60 & 42.15 \\ \hline
    \multicolumn{6}{l}{Period 2} \\ \hline
    ELM (30) & 55.18 & 47.15 & 56.38 & 42.76 & \textbf{1.80} \\
    ELM (300) & 54.23 & 52.57 & 52.95 & 49.54 & 20.81 \\
    LR & 55.02 & 44.95 & 56.46 & 41.78 & 41.65 \\ \hline
    \multicolumn{6}{l}{Period 3} \\ \hline
    ELM (30) & 56.19 & 48.22 & 59.06 & 44.63 & \textbf{1.78} \\
    ELM (300) & 54.66 & 53.06 & 56.16 & 52.58 & 30.19 \\
    LR & 55.79 & 45.48 & 59.00 & 44.13 & 42.03 \\ \hline
    \multicolumn{6}{l}{Whole Period} \\ \hline
    ELM (30) & 56.08 & 47.41 & 58.61 & 44.94 & \textbf{4.05} \\
    ELM (300) & 55.32 & 52.64 & 57.14 & 51.76 & 41.48 \\
    LR & 55.92 & 45.27 & 58.48 & 43.86 & 88.89 \\ \hline
    \end{tabular}
    \label{predict}
\end{table}
\begin{figure}[H]
    \centering
    \includegraphics[width=0.75\linewidth]{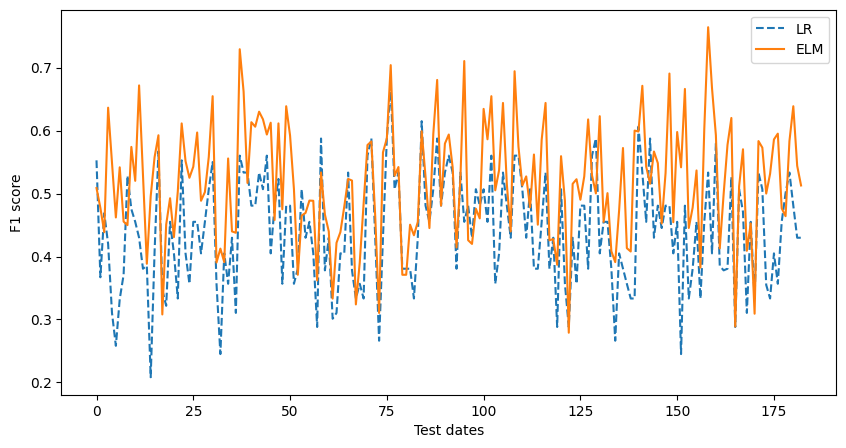}
    \caption{Out-of-sample classification F1 score of ELM ($L = 300$) and LR. The test period is from 2022-04-01 to 2022-12-31.}
    \label{accuracy}
\end{figure}

LR is a fundamental supervised learning algorithm used primarily for binary classification tasks, where the goal is to predict one of two possible outcomes. Unlike linear regression, which predicts continuous values, logistic regression predicts probabilities using the logistic function, mapping outputs to a range between 0 and 1.

Table \ref{predict} compares the predictive capabilities of ELM and LR under fixed architectural parameters ($L = 30$ hidden nodes, scale parameter = 0.01). The results demonstrate ELM's superior training accuracy relative to LR, while maintaining at least equivalent out-of-sample performance. This sustained advantage is further corroborated by Figure \ref{accuracy}, which visualizes daily classification accuracy throughout the test period. From the figure, ELM achieves better classification accuracy in average, while requiring significantly less computational time for model training.

The results can be justified by the network structure of ELM, which generates more complicated interactions between variables. In addition, LR requires iterative algorithms to converge to a solution while ELM directly estimates its network parameters, which makes ELM much faster for small neuron numbers.


\subsection{Completing IVS}
This experiment evaluates ELM for reconstructing implied volatility surfaces (IVS) using S\&P 500 options data sourced from OptionMetrics (January 4 – February 28, 2023). Specifically, we learn the mapping from option moneyness and maturity to the corresponding implied volatility. To train the mapping, we follow the following procedure:
\begin{enumerate}[itemsep=1pt,parsep=2pt,topsep=2pt]
    \item For each time to maturity, we obtain the risk-free rate from interpolating the treasury bill rates on the given date.
    \item Obtain option log moneyness from $k \equiv \frac{K}{F_t} = \frac{K}{S_t}e^{rT}.$
    \item Learn the mapping $(T, k) \mapsto \operatorname{IV}$ by an ELM/GPR framework.
\end{enumerate}

To reduce the data noise, we clean the data according to the following criteria (following the standard procedure in \cite{bardgett2019inferring}). We remove observations that
    \begin{itemize}[itemsep=1pt,parsep=2pt,topsep=2pt]
    \item are in-the-money ($k > 0$ for put options and $k < 0$ for call options) 
    \item violate of standard arbitrage conditions
        \item have log moneyness outside $[-1.2, 0.3]$
        \item exceed 3.0 years to maturity
        \item have missing implied volatility values
    \end{itemize}
After cleaning, the dataset contains an average of 2,658 daily implied volatility points, whose distributions are displayed in Appendix \ref{Dist}. We then partition the dataset into 80\% training and 20\% testing subsets via random sampling.

For benchmark comparison, GPR is implemented under the design that aligns with \cite{de2018machine} with kernel parameters optimized through grid-based searching. The ELM configuration utilizes a hyperbolic tangent activation function, 1,000 hidden nodes ($L = 1000$), and a fixed scale parameter $\sigma = 0.5$. We highlight the methodological difference that no parameter is tuned in ELM during the training process. 
\begin{figure}[H]
	\centering
	\begin{minipage}{0.49\linewidth}
\vspace{3pt}\centerline{\includegraphics[width=\textwidth]{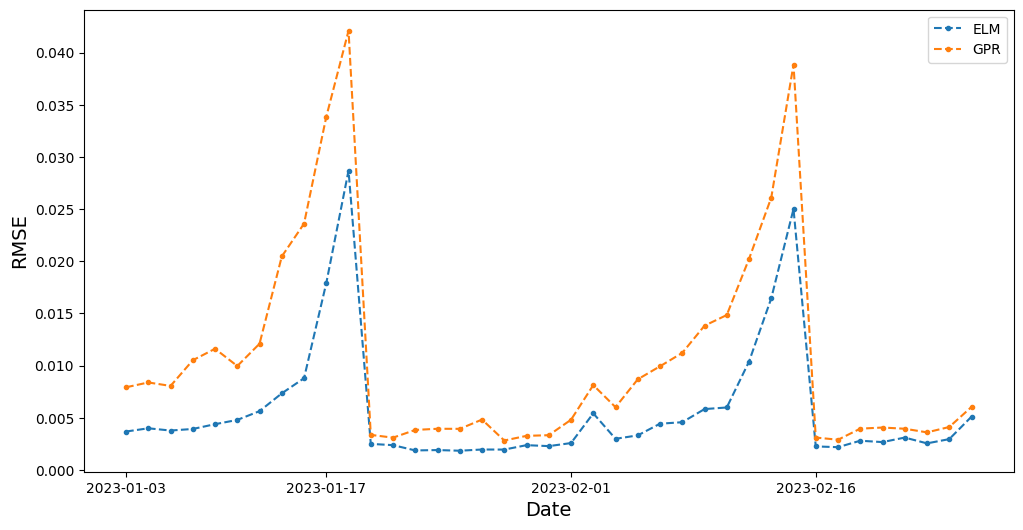}}
		\centerline{}
	\end{minipage}
	\begin{minipage}{0.49\linewidth}
		\vspace{3pt}
\centerline{\includegraphics[width=\textwidth]{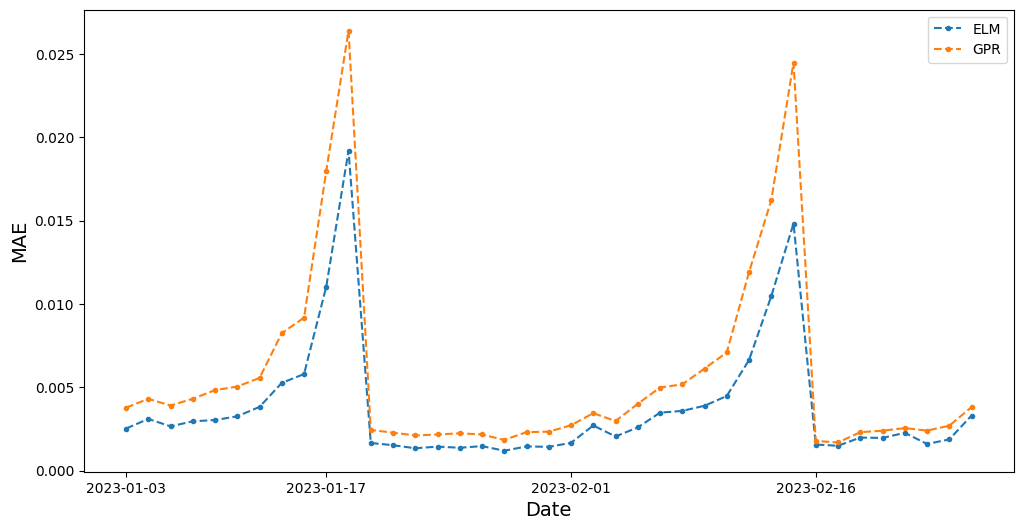}}
	 
		\centerline{}
	\end{minipage}
	\caption{The RMSE (left) and MAE (right) of ELM and GPR in the test sample. The ELM uses a hyperbolic tangent activation function with $L = 1000$ and $\sigma = 0.5$. The GPR configuration aligns with \cite{de2018machine} with kernel parameters optimized through grid-based searching.}
	\label{IVS}
\end{figure}

Figure \ref{IVS} compares the out-of-sample predictive accuracy of ELM and GPR in implied volatility surface completion. Despite operating without hyperparameter optimization, ELM systematically outperforms GPR across all testing dates throughout the evaluation period. 


After completing the IVS for each date, we validate the non-arbitrage condition of the trained model through the following numerical procedure. 
\begin{description}[itemsep=1pt,parsep=2pt,topsep=2pt]
    \item[Step 1] Create a meshgrid of $\{(T_{i}, K) \mid i=1, \cdots, N\}$ based on the boundaries of the original input data.
    \item[Step 2] Compute model IVS by applying the pre-trained ELM on the meshgrid.
    \item[Step 3] Assume the current price of the underlying $S = 1$, compute the call prices by applying the Black-Scholes formula.
    \item[Step 4] Validate the absence of arbitrage by checking the following conditions for $C(K,T)$ defined on $K\in [\min{K_i}, \max{K_i}]$ and $T\in [\min{T_i}, \max{T_i}]$ (\cite{fengler2009arbitrage}):
    \begin{enumerate}[itemsep=1pt,parsep=2pt,topsep=2pt]
        \item $C(K, \cdot)$ is non-decreasing for $K\in [\min{K_i}, \max{K_i}]$.
        \item $C(\cdot, T)$ is non-increasing for $T\in [\min{T_i}, \max{T_i}]$.
        \item $C(\cdot, T)$ is a convex function for $T\in [\min{T_i}, \max{T_i}]$.
    \end{enumerate}
\end{description}

We examine the non-arbitrage conditions on the IVS of range $K \equiv e^k\in [0.7, 1.2]$ and $T\in [0.05, 1]$. The IVS range is discretized evenly into a meshgrid $100 \times 100$. We define the violation rate of arbitrage of type one as the percentage of moneynesses under which $C(K, \cdot)$ is not non-decreasing. The violation rates of the other two types are defined likewise.

\begin{figure}[H]
	\centering
	\begin{minipage}{0.32\linewidth}
\vspace{3pt}\centerline{\includegraphics[width=\textwidth]{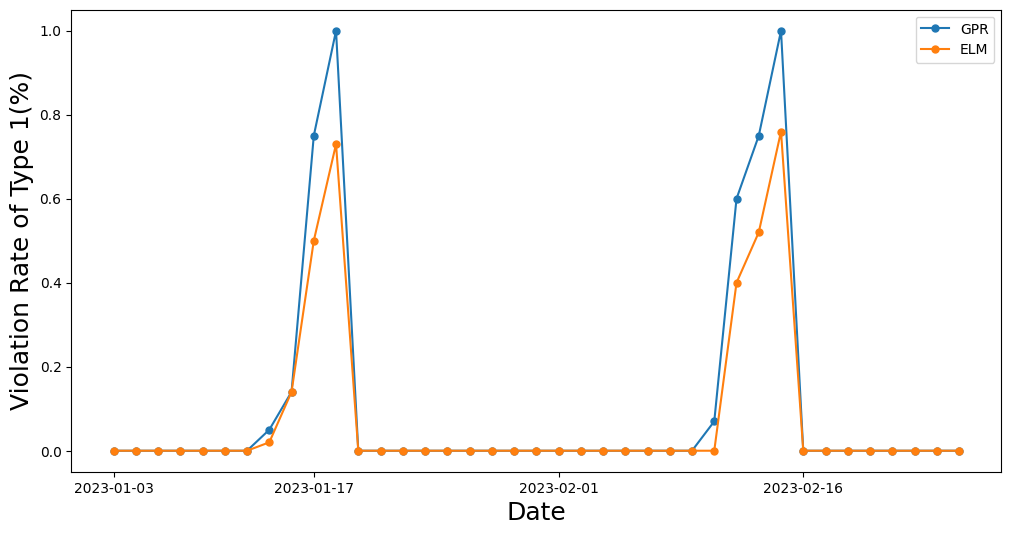}}
		\centerline{}
	\end{minipage}
	\begin{minipage}{0.32\linewidth}
		\vspace{3pt}
\centerline{\includegraphics[width=\textwidth]{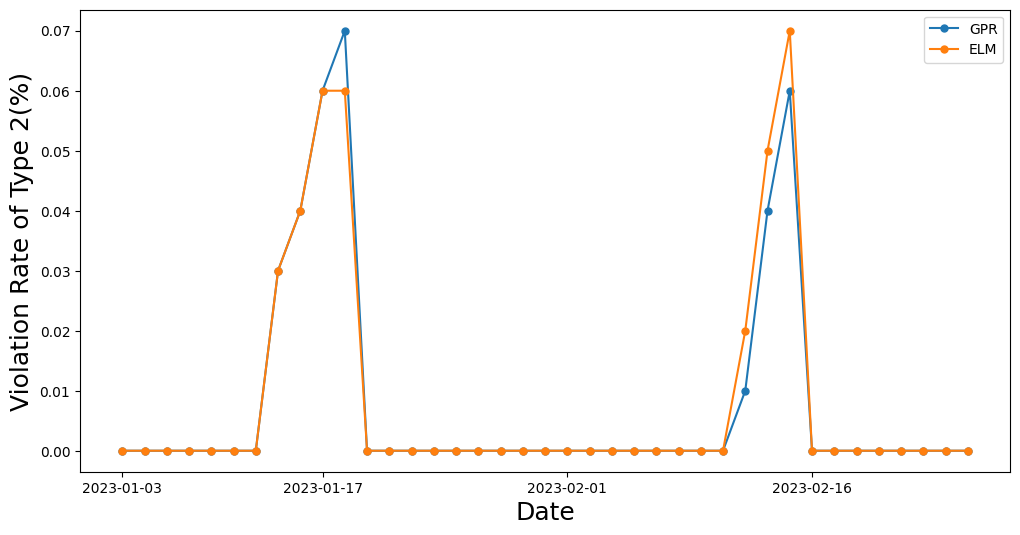}}
	 
		\centerline{}
	\end{minipage}
    	\begin{minipage}{0.32\linewidth}
		\vspace{3pt}
\centerline{\includegraphics[width=\textwidth]{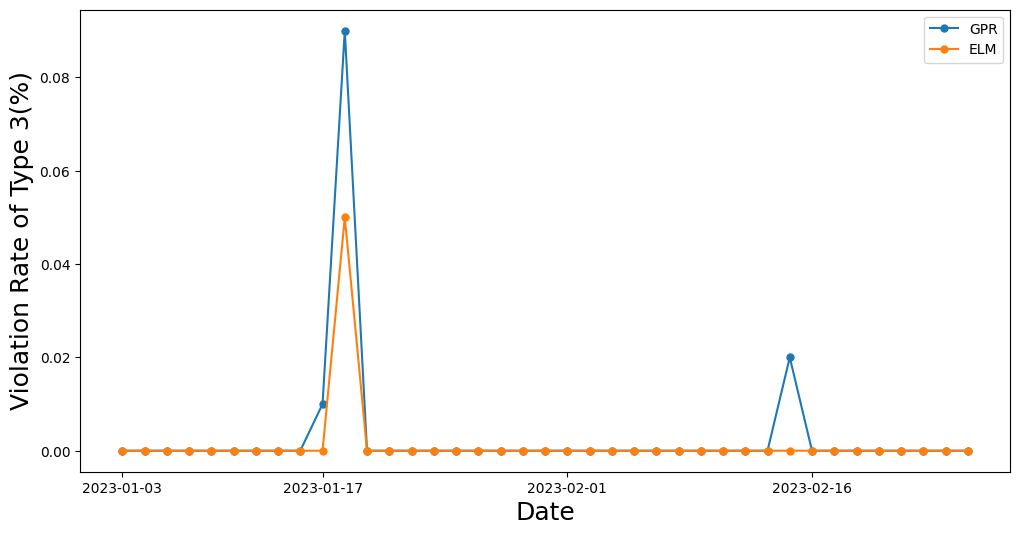}}
	 
		\centerline{}
	\end{minipage}
	\caption{Violation rates of the monotonicity condition of $T$ (left), $K$ (middle) and the convexity condition (right) in ELM and GPR method. The violation rate is defined as the percentage of moneynesses (maturities) where the non-decreasing (convexity) condition is not satisfied.}
	\label{vio}
\end{figure}

The numerical tests in Figure \ref{vio} show that both methods admit no arbitrage opportunities on most dates of the two-month period, but admit some arbitrage of the three types on certain dates that correspond to the relatively high out-of-sample error test shown in Figure \ref{IVS}. In general, ELM-generated IVS admits fewer violations than GPR, more prominently in terms of the convexity condition. 

\begin{figure}[H]
	\centering
	\begin{minipage}{0.32\linewidth}
\vspace{3pt}\centerline{\includegraphics[width=\textwidth]{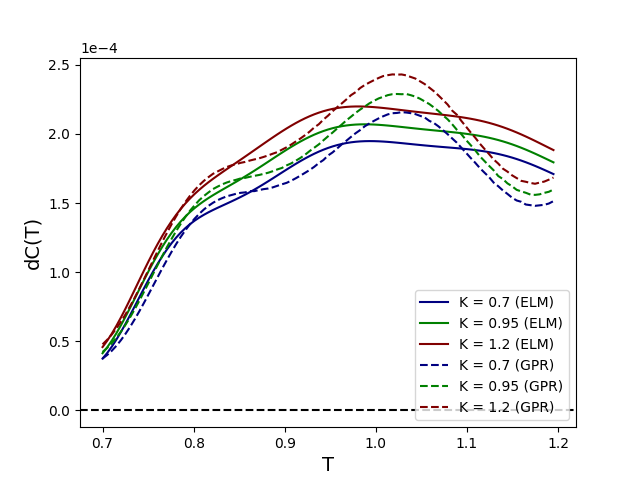}}
		\centerline{}
	\end{minipage}
	\begin{minipage}{0.32\linewidth}
		\vspace{3pt}
\centerline{\includegraphics[width=\textwidth]{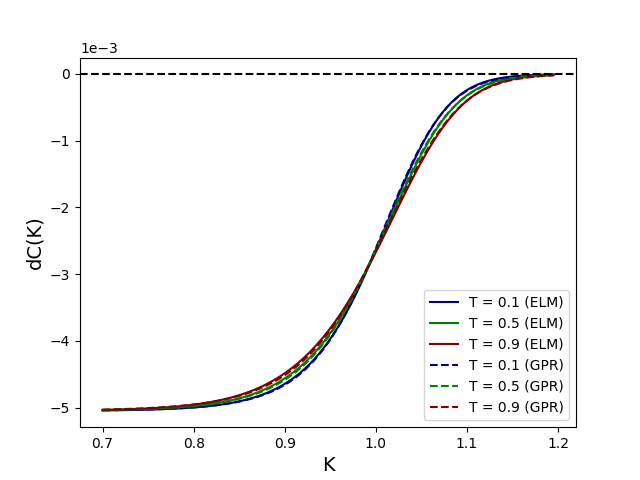}}
	 
		\centerline{}
	\end{minipage}
    	\begin{minipage}{0.32\linewidth}
		\vspace{3pt}
\centerline{\includegraphics[width=\textwidth]{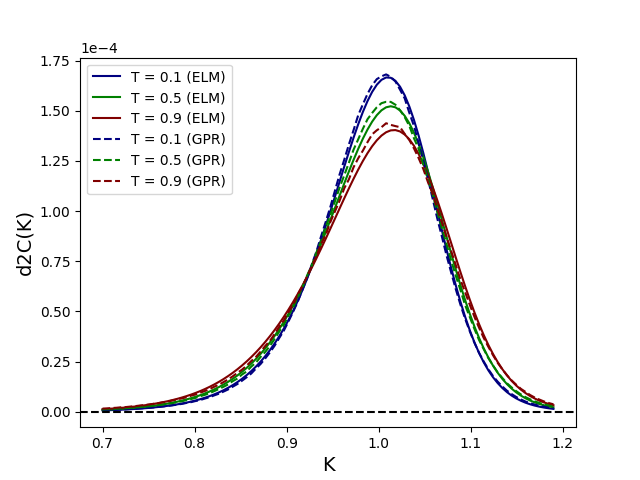}}
	 
		\centerline{}
	\end{minipage}

	\begin{minipage}{0.32\linewidth}
\vspace{3pt}\centerline{\includegraphics[width=\textwidth]{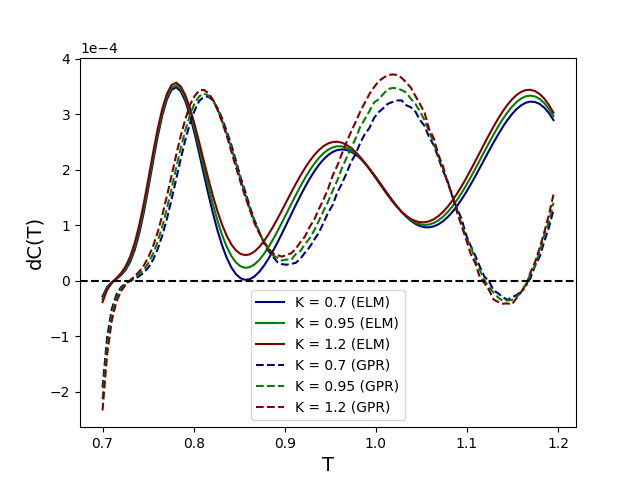}}
		\centerline{}
	\end{minipage}
	\begin{minipage}{0.32\linewidth}
		\vspace{3pt}
\centerline{\includegraphics[width=\textwidth]{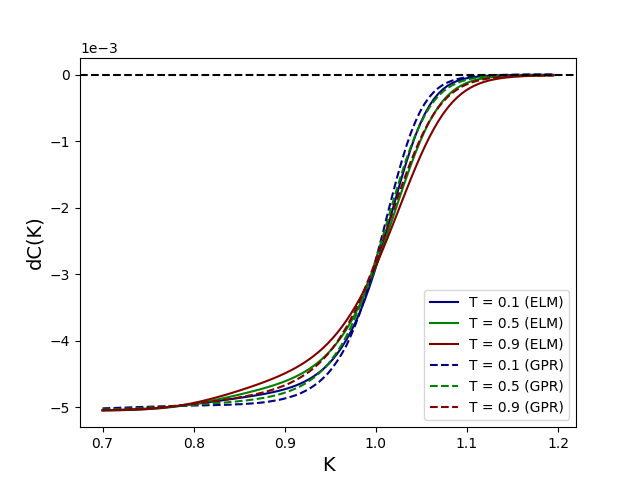}}
	 
		\centerline{}
	\end{minipage}
    	\begin{minipage}{0.32\linewidth}
		\vspace{3pt}
\centerline{\includegraphics[width=\textwidth]{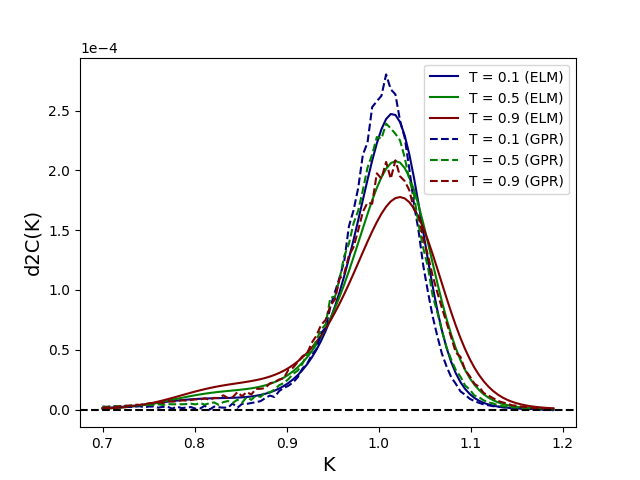}}
	 
		\centerline{}
	\end{minipage}
	\caption{Validation of the non-arbitrage conditions on 2023-01-03 (first row, no arbitrage observed) and on 2023-01-18 (second row, arbitrage observed). The x-axis is discretized into 100 points and the values represent the first-order difference $\mathrm{d}C(T)$, $\mathrm{d}C(K)$, and the second-order difference $\mathrm{d}^2C(K)$.}
	\label{conds}
\end{figure}

The disparity likely stems from fundamental methodological differences: GPR's propensity to overfit localized noise patterns amplifies surface irregularities, whereas ELM's generalization capacity – controlled through architectural parameters (node number and scale parameter) – enforces smoother representations. In fact, \cite{jacot2018neural} argues that the kernel in GPR acts similarly to the dynamics of an ANN in the infinite-width limit, which makes the training and generalization properties of wide ANNs analogous to those of GPR. Compared to ELM, whose structure is equivalent to a single-layer ANN, the interpolation capability of a multilayer equivalence is more likely to overfit. Figure \ref{conds} explores the disparity by comparing the monotonicity and convexity of $C(K, T)$ directly in both methods. We see from the left subplot that the ELM-generated IVS tends to be more robust with smoother temporal derivatives. In addition, the bumps generated by GPR observed in $\mathrm{d}^2C(K)$ are absent in ELM-generated IVS, which also demonstrates the superior IVS-completing performance of ELM.

\section{Applications in Unsupervised Learning}
\label{sec 4}

In this section, we apply ELM within an unsupervised learning framework to obtain numerical solutions to financial PDEs. Although the Black-Scholes model has known limitations, it remains a cornerstone of mathematical finance and in the industry. We use it here as a proof of concept to validate our numerical results, over its closed-form solutions for various financial derivatives. We begin by introducing the theory of Physics-Informed ELM, followed by numerical experiments on several derivative products, including European options, rainbow options, and barrier options.


Consider the following PDE system:
$$
\left\{\begin{aligned}
    &\frac{\partial}{\partial t}u(\overrightarrow{x},t) + \mathcal{L}u(\overrightarrow{x},t) = R(\overrightarrow{x},t), \quad (\overrightarrow{x},t) \in \Omega \times [0,T]\\
    & u(\overrightarrow{x},t) = B(\overrightarrow{x},t), \quad (\overrightarrow{x},t) \in \partial\Omega \times [0,T], \\
    & u(\overrightarrow{x},0) = F(\overrightarrow{x}), \quad \overrightarrow{x} \in \Omega,
\end{aligned}\right.$$
where \(\mathcal{L}\) is a linear differential operator and \(\partial\Omega\) is the boundary of computational domain \(\Omega\). \cite{dwivedi2020physics} proposed a Physics informed ELM to approximate $u(\mathbf{x}, t)$ by the output of ELM: $u(\overrightarrow{x},t) = \operatorname{ELM}(\mathbf{x}, t).$ Suppose that $\Omega \in \mathbf{R}^d$. Define $\mathbf{X} = [\mathbf{x}, t, 1]^\top$, $\mathbf{w}^X = [\mathbf{w}^X_1, \mathbf{w}^X_2, \dots, \mathbf{w}^X_L]^\top$, $\mathbf{w}^T = [w^T_1, w^T_2, \dots, w^T_L]^\top$ and $\mathbf{b} = [b_1, b_2, \dots, b_L].$ Under the weights and bias of the hidden layer given by $(\mathbf{w}^X, \mathbf{w}^T, \mathbf{b})$, the output of the $k$-th hidden neuron is then $$h_k = G(z_k) \quad k = 1, 2, \dots, L$$
for activation function $G(\cdot)$, where $z_k =  [\mathbf{w}^X_k, w^T_k, b_k]\cdot \mathbf{X}$. The ELM output is given by $$f(\mathbf{X}) = \mathbf{h}_{1\times k}\boldsymbol{\beta}_{k\times 1}.$$
We can derive from the operations in the ELM network that
$$\left\{\begin{aligned}
    \frac{\partial^p f_k}{\partial x^p_l} &= (w^X_{kl})^p \frac{\partial^p G}{\partial \mathbf{z}^p} \boldsymbol{\beta}, \; p = 1,2\\
    \frac{\partial f_k}{\partial t} & = w^T_k \frac{\partial G}{\partial \mathbf{z}} \boldsymbol{\beta},
\end{aligned}\right.$$
where $l = 1, 2, \dots, d$ and $k = 1, 2, \dots, L.$ The mixed partial derivatives can be deduced likewise.

In the training process, we randomly sample points $(\mathbf{x}, t)$ with number $N_f, N_{bc}, N_{ic}$ in the domain $\Omega \times [0, T]$, boundary $\partial \Omega\times [0, T]$ and $\Omega \times T$, respectively. We define the corresponding training errors as $\boldsymbol{\xi}_{f}$, $\boldsymbol{\xi}_{bc}$ and $\boldsymbol{\xi}_{ic}$ with
$$\left\{\begin{aligned}
    \boldsymbol{\xi}_f &= \frac{\partial f}{\partial t} + \mathcal{L}f - R, \quad (\mathbf{x}, t) \in \Omega \times [0, T],\\
    \boldsymbol{\xi}_{bc} &= f - B, \quad (\mathbf{x}, t) \in \partial \Omega \times [0, T],\\
    \boldsymbol{\xi}_{ic} &= f(\cdot, 0) - F, \quad \mathbf{x} \in \Omega.
\end{aligned}\right.$$

To solve the linear PDE, we need to mininize $ \| \boldsymbol{\xi}_f\|_2^2 + \| \boldsymbol{\xi}_{bc}\|_2^2 + \|\boldsymbol{\xi}_{ic}\|_2^2,$
which leads to a linear equation system represented as
$$\mathbf{H}\cdot \boldsymbol{\beta} = \mathbf{Y},$$
where $\mathbf{H} \in \mathbf{R}^{N \times {L}}$, with $N = N_{f} + N_{bc} + N_{ic}$, is the hidden matrix determined by the linear operator and the network configuration, and $\mathbf{Y}$ depends on functions $R$, $B$ and $F$ given by the PDE. We generally set $N > L$ to ensure a sufficiently large training set.

Finally, the solution of the linear equation system under least squared error is given by Moore–Penrose generalized inverse: $\boldsymbol{\beta} = (\mathbf{H}^\top \mathbf{H} + C\mathbf{I})^{-1}\mathbf{H}^\top \mathbf{Y},$ which is equivalent to the solution of the following optimization problem: $\arg \min_{\boldsymbol{\beta}} J$, where
$$J = \frac{1}{2}C \| \boldsymbol{\beta} \|^2 + \frac{1}{2} \left( \frac{\boldsymbol{\xi}_f^\top \boldsymbol{\xi}_f}{N_f} + \frac{\boldsymbol{\xi}_{bc}^\top \boldsymbol{\xi}_{bc}}{N_{bc}} + \frac{\boldsymbol{\xi}_{ic}^\top\boldsymbol{\xi}_{ic}}{N_{ic}} \right),$$
with the regularization parameter $C.$ Increasing $C$ can enhance the generalization ability of ELM.

Next, we apply the theory of physics informed ELM to solve the price of exotic options whose terminal condition, instead of initial condition, is known. The call price of a European option is
$$V(t, \mathbf{u}) = e^{-r(T-t)} \mathbb{E}[H(\mathbf{S}_T) \mid \mathbf{S}_t = e^\mathbf{u}],$$
which is the unique viscosity solution of the partial differential equation
$$\frac{\partial V}{\partial t}(t, \mathbf{u}) + \frac{1}{2} \sum_{1 \le i, j\le m} \rho_{ij}\sigma_i\sigma_j\frac{\partial^2 V}{\partial u_{i}\partial _j}(t, \mathbf{u})  + r \sum_{l=1}^m\frac{\partial V}{\partial u_l} (t,\mathbf{u}) - rV(t,\mathbf{u})= 0, \; (t,\mathbf{u}) \in (0, T) \times \mathbb{R}^m,$$
with terminal condition $V(T, \mathbf{u}) = H(\mathbf{u})$ and $u_l = \ln S_l, \; l = 1, \dots, m.$ 

When $m = 1$ and $H(u) = (K - e^u)^+$, the put option admits the analytic price
$$V(t, u) =  e^u \mathcal{N}(d_1) - Ke^{-r(T-t)}\mathcal{N}(d_2),$$
where $$d_2 = \frac{1}{\sigma \sqrt{T}}\left(u - \ln (K)+ (r - \frac{\sigma^2}{2})T\right)$$ and $d_1 = d_2 + \sigma\sqrt{T}.$

\begin{figure}[H]
	\centering
    	\begin{minipage}{0.45\linewidth}
\vspace{3pt}\centerline{\includegraphics[width=\textwidth]{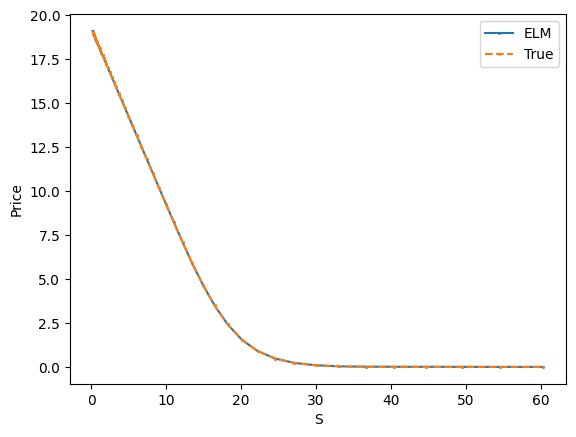}}
		\centerline{}
	\end{minipage}
	\begin{minipage}{0.45\linewidth}
		\vspace{3pt}
\centerline{\includegraphics[width=\textwidth]{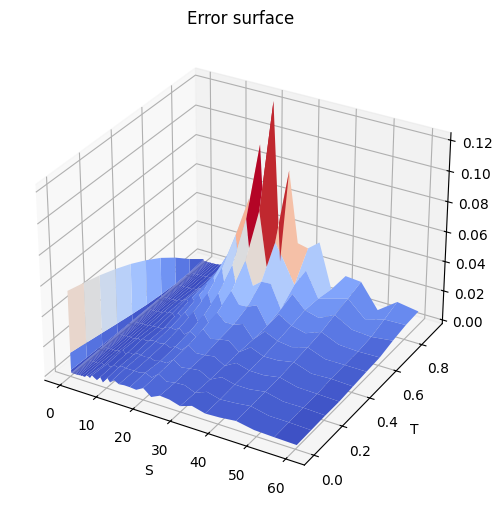}}
	 
		\centerline{}
	\end{minipage}
    	\begin{minipage}{0.45\linewidth}
\vspace{3pt}\centerline{\includegraphics[width=\textwidth]{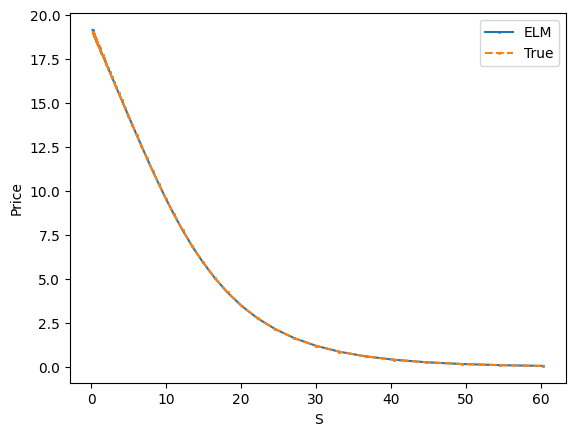}}
		\centerline{}
	\end{minipage}
	\begin{minipage}{0.45\linewidth}
		\vspace{3pt}
\centerline{\includegraphics[width=\textwidth]{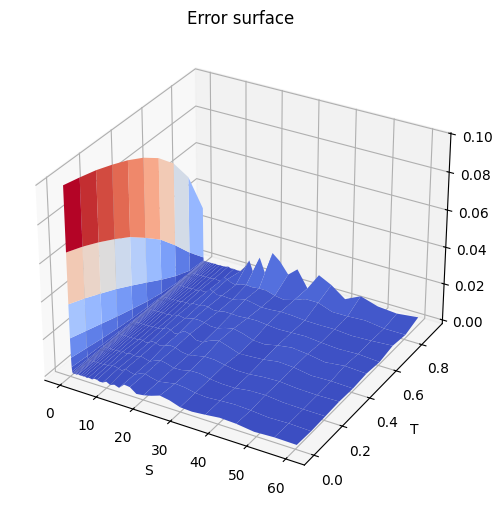}}
	 
		\centerline{}
	\end{minipage}
	\caption{Comparison of ELM predictions and the analytical prices for put options. We set $r = 0.04$, $T= 1$ with hidden node number $5000$ and scale $1$. The hyperbolic tangent function, as chosen in \cite{dwivedi2020physics}, is adopted as the activation function. The left column compares the predictions and the Black-Scholes prices under $K = 15$, with the relative errors 0.00059 ($\sigma = 0.25$, first row) and 0.00143 ($\sigma = 0.5$, second row). And the right column shows the $L^1$ option price errors across different maturities and initial prices, with mean relative errors 0.00076 (first row) and 0.00103 (second row).}
	\label{num1}
\end{figure}

We further define the relative error as $$\text{RE} = \frac{\|V_{\text{True}} - V_{\text{ELM}}\|_2}{\|V_{\text{True}\|_2}}.$$ Figure \ref{num1} shows the performance of ELM in put options for $m = 1$. The relative error is 0.00059 for $\sigma = 0.25$ and 0.00143 for $\sigma = 0.5$. We also calculate on the implied volatility surface ($T \in [0, 1)$, $S \in [0, 60]$) a mean relative error of 0.00076 for $\sigma = 0.25$ and 0.00103 for $\sigma = 0.5$.

\begin{figure}[H]
	\centering
		\begin{minipage}[t]{0.45\linewidth}
		\centerline{\includegraphics[width=\textwidth]{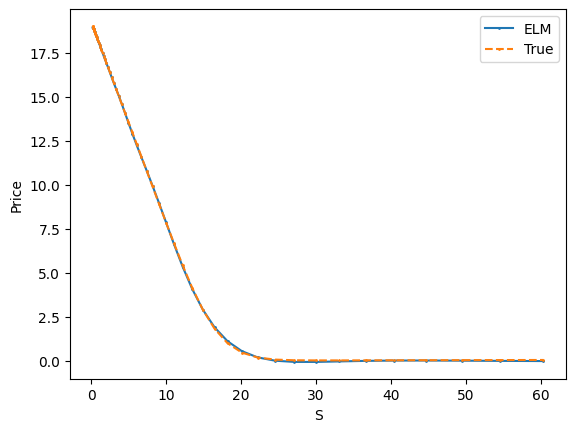}}
	\end{minipage}
	\begin{minipage}[t]{0.45\linewidth}
		\centerline{\includegraphics[width=\textwidth]{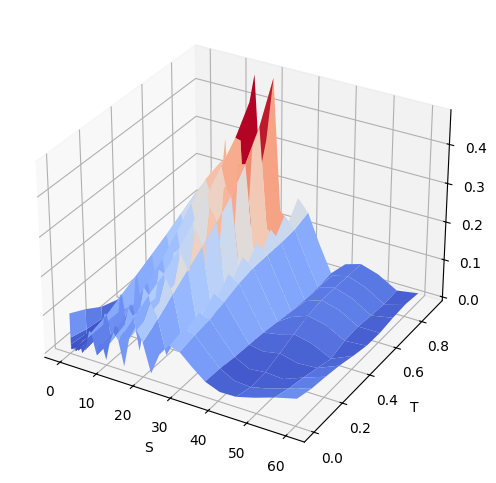}}
	\end{minipage}
	
	 \vspace{2pt} 
	
	\begin{minipage}[t]{0.45\linewidth}\vspace{-5pt}\centerline{\includegraphics[width=\textwidth]{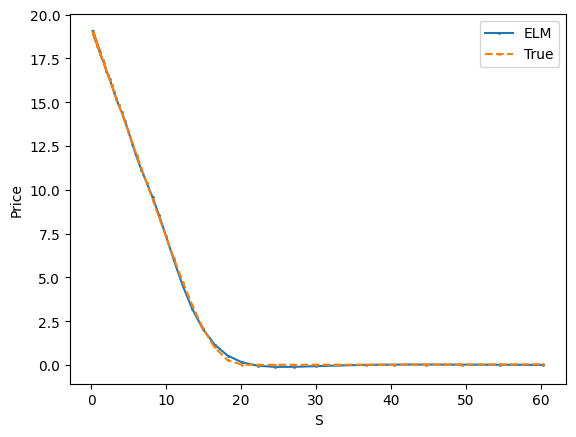}}
	\end{minipage}
	\begin{minipage}[t]{0.45\linewidth}
		\vspace{-5pt} \centerline{\includegraphics[width=\textwidth]{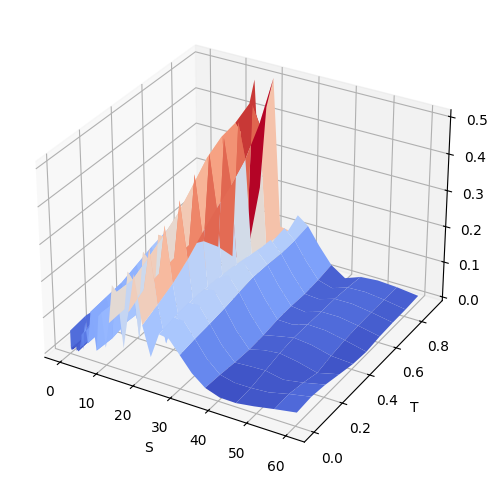}}
	\end{minipage}
	\caption{Comparison of ELM predictions and the analytical prices for rainbow put options. We set $r = 0.04$, $\sigma = 0.25$, $T= 1$ with hidden node number $5000$ and scale $1$. The left column compares the predictions and the Black-Scholes prices under $K = 20$, with the relative errors 0.00294 for $\rho = 0$ (first row) and 0.00545 for $\rho = -0.95$ (second row). And the right column shows the $L^1$ option price errors across different maturities and initial prices, with mean relative errors 0.00616 for $\rho = 0$ and 0.00776 for $\rho = -0.95$.}
	\label{num2}
\end{figure}

As an example of dimension $m = 2$, given $H(\ln\mathbf{S}_T) = \max\{K - \max\{S_{1,T}, S_{2, T}\}, 0\}$, the payoff of a put rainbow option on maximum of two assets, we examine how ELM produces the price of a rainbow option. The analytical price formula of rainbow options can be found in \cite{stulz1982options}. Figure \ref{num2} shows that the relative error at $T = 1$ is $0.00294$ for $\rho = 0$ and $0.00545$ for $\rho = -0.95$. The performance is weaker than the 1-dimensional European options, but is still good enough in practice.

We also explore the application of ELM in barrier options. We consider a double-barrier call option whose price satisfies the following PDE system:

\begin{equation}
    \left\{\begin{aligned}
    & \frac{\partial V}{\partial t}(t, u) + \frac{\sigma^2}{2} \sum_{l = 1}^m\frac{\partial^2 V}{\partial u_{l}^2}(t, u)  + r \sum_{l=1}^m\frac{\partial V}{\partial u_l} (t,u) - rV(t,u)= 0, \quad (t,u) \in (0, T) \times [L, U]\\
    & V(t, L) = 0, \; V(t, U) = 0, t\in [0, T]\\
    & V(T, u) = (e^u - K)^+, u\in (L, U).
    \end{aligned}\right.
\end{equation}
where $L, U$ are the knock-out bounds for the log price. Double-barrier options also admit analytic prices, as shown in Appendix \ref{Barrier}.

\begin{figure}[H]
	\centering
	\begin{minipage}{0.49\linewidth}
\vspace{3pt}\centerline{\includegraphics[width=\textwidth]{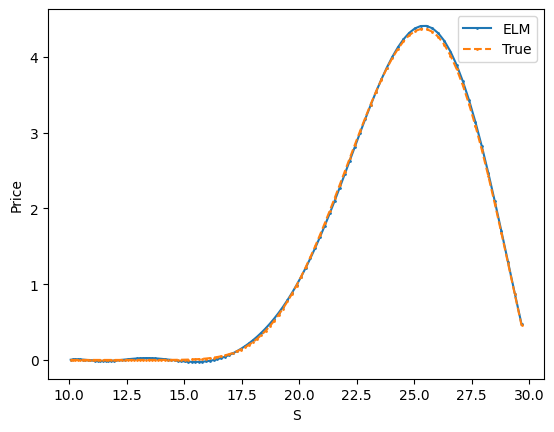}}
		\centerline{}
	\end{minipage}
	\begin{minipage}{0.49\linewidth}
		\vspace{3pt}
\centerline{\includegraphics[width=\textwidth]{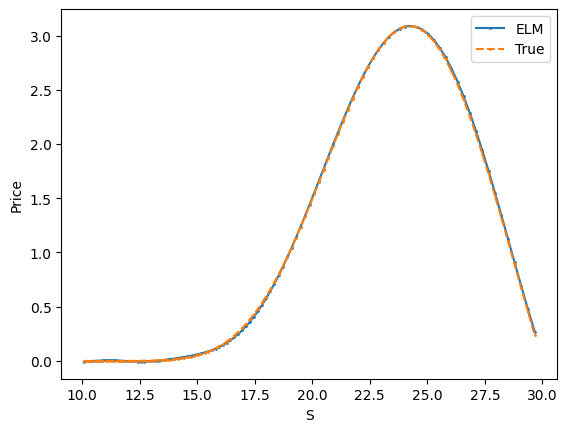}}
	 
		\centerline{}
	\end{minipage}
	\caption{Comparison of ELM predictions and the analytical prices for knock-out double-barrier call options with lower and upper bound 10 and 30, respectively. We set $r = 0.04$, $\sigma = 0.15$, $K = 20$ with hidden node number $5000$ and scale $1$. The left (right) plot compares the predictions and the Black-Scholes prices under $T = 0.5$ ($T = 1$), with relative error 0.01343 (0.00909).}
	\label{num3}
\end{figure}

\begin{figure}[H]
	\centering
	\begin{minipage}{0.49\linewidth}
\vspace{3pt}\centerline{\includegraphics[width=\textwidth]{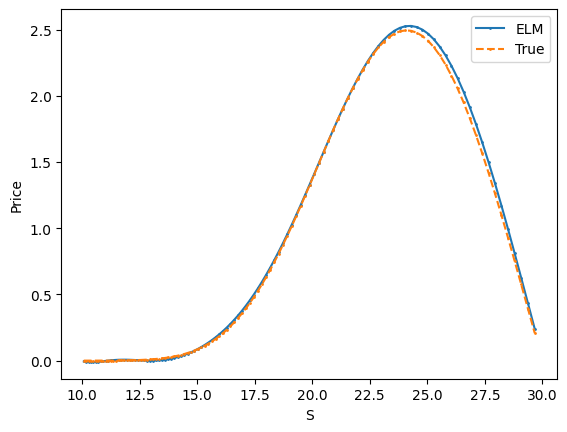}}
		\centerline{}
	\end{minipage}
	\begin{minipage}{0.49\linewidth}
		\vspace{3pt}
\centerline{\includegraphics[width=\textwidth]{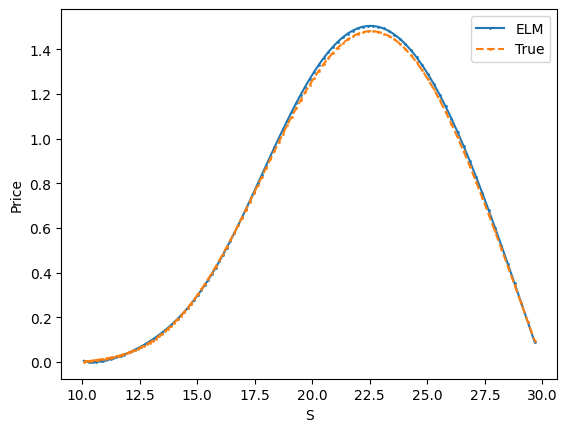}}
	 
		\centerline{}
	\end{minipage}
	\caption{Comparison of ELM solutions and the analytical prices for knock-out double-barrier call options with lower and upper bound 10 and 30, respectively. We set $r = 0.04$, $\sigma = 0.25$, $K = 20$ with hidden node number $5000$ and scale $1$. The left (right) plot compares the predictions and the Black-Scholes prices under $T = 0.5$ ($T = 1$), with relative error 0.02489 (0.01831).}
	\label{num4}
\end{figure}

As $\tau \to 0$, the price of the knock-out double-barrier call option becomes volatile for $S$ close to the upper bound. The option is more likely to be knocked out for large $S$, but also possibly pay off better for such initial prices. This sharp fall of the price around termination can result in greater discrepancy between ELM's predictions and the true prices.

\begin{remark}

Compared with classical PINNs, ELM runs more rapidly since it formulates a PDE into a linear equation system and solves the equations via efficient convex optimization. The relevant speed comparison was shown in \cite{dwivedi2020physics} that it takes ELM a few seconds to solve a linear PDE where DNN takes hours to achieve a comparable accuracy. We also note that ELM shows promising ability in overcoming the curse of dimensionality for learning solutions to certain financial PDEs, which is further elaborated in \cite{gonon2023random}. 
\end{remark}

\section{Conclusions}
\label{sec 5}
This paper has investigated the application of extreme learning machines in quantitative finance, demonstrating their effectiveness across supervised and unsupervised learning. Our results establish ELM as a computationally efficient alternative to traditional methods with respect to training and inference procedures while maintaining competitive accuracy.

In supervised learning tasks, ELM excels in three critical scenarios. First, for learning pricing functions in stochastic volatility models, ELM achieves training speed orders of magnitude faster than DNNs and GPR. This efficiency advantage persists even when comparing the inference speed, with ELM several times faster while maintaining comparable pricing accuracy. In addition, the EIR-ELM variant is also shown to effectively reduce model complexity without sacrificing accuracy in the task of learning pricing functions. Second, in high-frequency stock return forecasting, ELM classifiers outperform logistic regression in both accuracy and training speed (20× faster), which demonstrates its advantages for intraday trading data analysis. Third, for constructing implied volatility surfaces, ELM generates arbitrage-free surfaces with smoother derivatives and fewer arbitrage violations than GPR, making it more robust and practical in applications.

For unsupervised PDE solving, ELM overcomes key limitations of DNN-based approaches. In terms of training efficiency, ELM-based PINNs solve Black-Scholes-type equations orders-of-magnitude faster than traditional DNN implementations, which results from its formation of the problem as a linear equation system. Regarding high-dimensional PDEs, ELM is also shown to obtain promising accuracy.

Two promising future discussions emerge from this work. First, extending ELM-based PINNs to other path-dependent derivatives, such as American options and credit derivatives with early exercise features, remains an open challenge. Second, integrating ELM with online learning frameworks could enable real-time recalibration in non-stationary markets, further enhancing their practicality in trading and risk management.

In summary, this work positions ELM as a versatile tool for quantitative finance that combines the speed of traditional numerical methods with the flexibility of machine learning in supervised and unsupervised learning, which makes ELM potentially compelling for high-frequency trading systems, real-time risk management, and multi-asset derivative pricing.

\begin{appendices}
\section{Feature Descriptions}
\label{A}
In the task of intraday stock price movement prediction, we construct 12 features features. We introduce the following notation to define the features:
\begin{itemize}[itemsep=1pt,parsep=2pt,topsep=2pt]
    \item \( t_i \): the timestamp of the beginning of the \( i \)-th interval in the baseline execution strategy.
    \item \( u_i \): the timestamp of the \( i \)-th trade.
    \item \( P_i \): the transaction price of the \( i \)-th trade.
    \item \( q_i \): the transaction volume of the \( i \)-th trade.
    \item \( P_t \): the stock price at time \( t \), defined as the price of the most recent trade before or at time \( t \):
    
    \[
    P_t = P_{\max\{i:u_i \leq t\}}.
    \]
    
    \item \(\mathcal{P}_{t,k}^{Bid}\): the set of the top \( k \) bid prices in the order book at time \( t \).
    \item \(\mathcal{P}_{t,k}^{Ask}\): the set of the top \( k \) ask prices in the order book at time \( t \).
    \item \( Q_{t,P} \): the volume of limit orders placed at price \( P \) in the order book at time \( t \).
    
    \item \( OI_{t,k} \): the order imbalance at the \( k \)-th level at time \( t \), that is
    
    \[
    OI_{t,k} = \frac{Q_{t,k}^{Bid} - Q_{t,k}^{Ask}}{Q_{t,k}^{Bid} + Q_{t,k}^{Ask}},
    \]
    
    where
    
    \[
    Q_{t,k}^{Bid} = \sum_{P \in \mathcal{P}_{t,k}^{Bid}} Q_{t,P}, \quad Q_{t,k}^{Ask} = \sum_{P \in \mathcal{P}_{t,k}^{Ask}} Q_{t,P}.
    \]
\end{itemize}

To predict the stock price at $t_{i+1},$ we consider two categories of features: features
derived from historical transactions and features derived from the limit order book. In the following, we provide the framework for these features.
 
 The features derived from historical transactions include:
 \begin{itemize}[itemsep=1pt,parsep=2pt,topsep=2pt]
    \item \textbf{Open Price}: The opening price of the time interval \([t_{i-1},t_{i})\). Open Price = \(P_{t_{i-1}}\).
    
    \item \textbf{Close Price}: The closing price of the time interval \([t_{i-1},t_{i})\). Close Price = \(\lim\limits_{t\to t_{i}-0}P_{t}\).
    
    \item \textbf{High Price}: The highest price during the time interval \([t_{i-1},t_{i})\). High Price = \(\max\limits_{t\in[t_{i-1},t_{i})}P_{t}\).
    
    \item \textbf{Low Price}: The lowest price during the time interval \([t_{i-1},t_{i})\). Low Price = \(\min\limits_{t\in[t_{i-1},t_{i})}P_{t}\).
    
    \item \textbf{Volume-Weighted Average Price (VWAP)}: 
    \[
    \text{VWAP} = \left\{\begin{array}{ll}
    \frac{\sum\limits_{u_{i}\in[t_{i-1},t_{i})}P_{i}q_{i}}{\sum\limits_{u_{i}\in[t_{i-1},t_{i})}q_{i}}, \quad \sum\limits_{u_{i}\in[t_{i-1},t_{i})}q_{i}>0, \\
    \text{Close Price, otherwise.}
    \end{array}\right.
    \]
    
    \item \textbf{Time-Weighted Average Price (TWAP)}: 
    \[
    \text{TWAP} = \frac{\int_{t_{i-1}}^{t_{i}} P_t \, dt}{t_{i}-t_{i-1}}.
    \]
    
    \item \textbf{Market Trading Volume}: The total market trading volume during the time interval \([t_{i-1},t_{i})\). Trade Volume = \(\sum\limits_{u_{i}\in[t_{i-1},t_{i})}q_{i}\).
\end{itemize}
The features derived from the limit order book include:
\begin{itemize}[itemsep=1pt,parsep=2pt,topsep=2pt]
    \item \textbf{Order Imbalance at \(t_{i}\):} The order imbalance for the top \(k\) buy and sell prices at time \(t_{i}\). Order Imbalance\({}_{t_{i},k}\) = OI\({}_{t_{i},k}\), where \(k=1,5,\infty\).
    
    \item \textbf{Time-Weighted Average Order Imbalance:} The time-weighted average order imbalance for the top \(k\) buy and sell prices over the past \(\Delta\) period. 
    \[
    \text{TWA Order Imbalance}_{t_{i},k}(\Delta) = \frac{\int_{t_{i}-\Delta}^{t_{i}} \text{OI}_{t,k} \, dt}{\Delta},
    \]
    where \(k=1,5,\infty\) and \(\Delta= \text{5min}\).
\end{itemize}
\section{A Brief Introduction of GPR}
\label{GPR}
Gaussian process regression is a non-parametric, Bayesian approach to regression that models a function \( f(\mathbf{x}) \) as a distribution over possible functions. It is fully defined by a mean function \( m(\mathbf{x}) \) and a covariance (kernel) function \( k(\mathbf{x}, \mathbf{x}') \).

\begin{enumerate}[itemsep=1pt,parsep=2pt,topsep=2pt]
    \item  Prior Distribution:  
   \[
   f(\mathbf{x}) \sim \mathcal{GP}\big(m(\mathbf{x}),\, k(\mathbf{x}, \mathbf{x}')\big),
   \]  
   where\( m(\mathbf{x}) \) is often assumed zero for simplicity and \( k(\mathbf{x}, \mathbf{x}') \) encodes similarity between inputs (e.g., RBF kernel).  

\item RBF Kernel Example:  
   \[
   k(\mathbf{x}, \mathbf{x}') = \sigma_f^2 \exp\left(-\frac{\|\mathbf{x} - \mathbf{x}'\|^2}{2\ell^2}\right),
   \]  
where \( \sigma_f^2 \), $\ell$ are parameters representing signal variance and length scale, respectively.  

\item Posterior Prediction:  
   Given training data \( \mathbf{X} \) (inputs) and \( \mathbf{y} \) (outputs), the predictive distribution for a new input \( \mathbf{x}_* \) is Gaussian:  
   \[
   p(f_* \mid \mathbf{X}, \mathbf{y}, \mathbf{x}_*) = \mathcal{N}(\bar{f}_*, \mathbb{V}[f_*]),
   \]  
from which we obtain the predictive mean:  
     \[
     \bar{f}_* = \mathbf{k}_*^T (\mathbf{K} + \sigma_n^2 \mathbf{I})^{-1} \mathbf{y}
     \]  
and predictive Variance:  
     \[
     \mathbb{V}[f_*] = k(\mathbf{x}_*, \mathbf{x}_*) - \mathbf{k}_*^T (\mathbf{K} + \sigma_n^2 \mathbf{I})^{-1} \mathbf{k}_*,
     \]  
where \( \mathbf{K} \) is the kernel matrix with the $(i,j)$ term \( \mathbf{k}_* := [k(\mathbf{x}_*, \mathbf{x}_1), \dots, k(\mathbf{x}_*, \mathbf{x}_N)]^\top \), \( K_{ij} = k(\mathbf{x}_i, \mathbf{x}_j) \), and \( \sigma_n^2 \) is the noise variance. 
\end{enumerate}

\section{Simplification of Network Structure Through EIR-ELM}
\label{EIR-Simp}
\begin{figure}[H]
    \centering
    \includegraphics[width=0.5\linewidth]{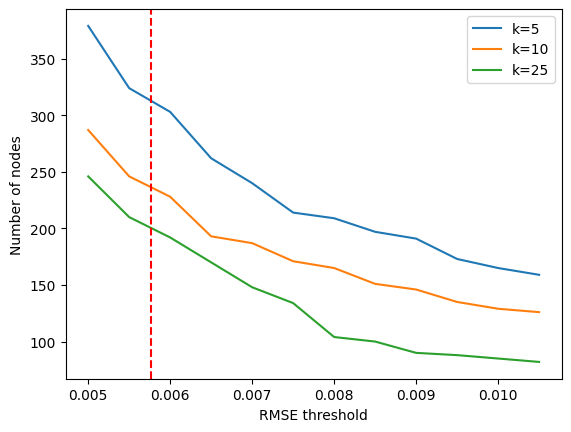}
    \caption{Simplification of neural network structure through EIR-ELM. The values correspond to the number of nodes needed to achieve the specified RMSE levels. A dataset size of 2000 is used. The red vertical line $x = 0.00577$ is the performance of GPR on the same dataset.}
    \label{EIR}
\end{figure}

In the task of learning the Heston pricing function, the EIR-ELM method attains GPR's performance (achieved via a 2,000-dimensional kernel matrix) using fewer than 350 nodes shown in Figure \ref{EIR}. Moreover, the network is further simplified as the selection degree $k$ increases. This optimized configuration reduces memory overhead while maintaining computational efficiency during inference phases.

\section{Description of Option Dataset}
\label{Dist}
\begin{table}[H]
    \centering
        \caption{A description of the dataset after cleaning.  Each row represents the number of options/average implied volatility across different maturity ranges. Each column represents the corresponding values accross different log moneyness ranges.}
\begin{tabular}{cccccc}
    \diagbox[width=5.5em, height=2em, innerleftsep=0pt, innerrightsep=0pt, dir=NW]{\footnotesize Log-mon}{\footnotesize Maturity} 
      & $\leq$7 days & 7–30 days & 30–90 days & 90–365 days & 1–3 years \\ \hline
    \multicolumn{6}{c}{Panel A: Number of Options} \\
    $[$-1.2, -0.9$]$   & 8     & 134   & 262    & 1114   & 459    \\
    (-0.9, -0.6]   & 26    & 194   & 417    & 2430   & 917    \\
    (-0.6, -0.3$]$   & 177   & 725   & 1421   & 8480   & 2934   \\
    (-0.3, 0$]$      & 1291  & 5769  & 11378  & 23659  & 5627   \\
    (0, 0.3$]$       & 792   & 4194  & 7435   & 16967  & 3940   \\ \hline
    \multicolumn{6}{c}{Panel B: Average Implied Volatility} \\
    $[$-1.2, -0.9$]$   & 2.28  & 1.38  & 0.82   & 0.56   & 0.42   \\
    (-0.9, -0.6$]$   & 2.08  & 1.02  & 0.65   & 0.46   & 0.37   \\
    (-0.6, -0.3$]$   & 1.50  & 0.63  & 0.44   & 0.34   & 0.30   \\
    (-0.3, 0$]$      & 0.56  & 0.31  & 0.25   & 0.24   & 0.23   \\
    (0, 0.3$]$       & 0.44  & 0.21  & 0.17   & 0.16   & 0.16   \\\hline
\end{tabular}
    \label{dist data}
\end{table}

\section{Formula for Double-barrier Options}
\label{Barrier}
It is shown in \cite{kunitomo1992pricing} that a double-barrier call option with upper (lower) bound $L$ ($U$) admits the following analytical price: 
$$
\begin{aligned}V(t, S; K, T, L, U) &= S \sum_{n=-\infty}^{\infty} \left\{ \left( \frac{F^n}{E^n} \right)^{c_{n}} [\mathcal{N}(d_{1,n}) - \mathcal{N}(d_{2,n})] - \left( \frac{E^{n+1}}{F^n S} \right)^{c_{n}} [\mathcal{N}(d_{3,n}) -\mathcal{N} (d_{4,n})] \right\}  \\
& - K e^{-r\tau} \sum_{n=-\infty}^{\infty} \left\{ \left( \frac{F^n}{E^n} \right)^{c_{n} - 2} \times [\mathcal{N}(d_{1,n} - \sigma \sqrt{\tau}) - \mathcal{N}(d_{2,n} - \sigma \sqrt{\tau})] \right.\\
& -\left. \left(\frac{E^{n+1}}{F^n S} \right)^{c_{n} - 2} \times [\mathcal{N}(d_{3,n} - \sigma \sqrt{\tau}) - \mathcal{N}(d_{4,n} - \sigma \sqrt{\tau})] \right\},
\end{aligned}$$
with $c_n = \frac{2|r|}{\sigma^2} + 1,$ $F = e^U$, $E = e^L$ and 
\[
d_{1,n} = \frac{\ln(S / K) +  2n (U - L)+ (r + \sigma^2/2)\tau}{\sigma \sqrt{\tau}},
\]
\[
d_{2,n} = \frac{\ln(S/U) + 2n(U - L) + (r + \sigma^2/2)\tau}{\sigma \sqrt{\tau}},
\]
\[
d_{3,n} = \frac{\ln(L^{2}/KS) - 2n(U- L)+ (r + \sigma^2/2)\tau}{\sigma \sqrt{\tau}},
\]
\[
d_{4,n} = \frac{\ln(L^{2}/US) - 2n(U-L) + (r + \sigma^2/2)\tau}{\sigma \sqrt{\tau}}.
\]
\end{appendices}
\end{document}